\documentclass[%
 reprint,
 superscriptaddress,
 amsmath,amssymb,
 aps,
 prb,
 floatfix,
]{revtex4-2}

\usepackage{siunitx}
\DeclareSIUnit{\dBm}{dBm}
\usepackage[none]{hyphenat}
\usepackage{graphicx}
\usepackage{dcolumn}
\usepackage{bm}
\usepackage[utf8]{inputenc}
\usepackage[T1]{fontenc}
\usepackage{mathptmx}
\usepackage{textcomp}
\usepackage{bigints}
\usepackage{amsmath}
\usepackage{wrapfig}
\usepackage{ragged2e}
\usepackage{enumitem}

\usepackage[switch]{lineno}

\usepackage[table,xcdraw,dvipsnames]{xcolor}
\usepackage{subcaption}
\usepackage{booktabs}
\usepackage{float}
\usepackage{color}
\usepackage{ulem}
\usepackage{multirow}
\usepackage{hyperref}

\begin{document}

\title{Brillouin Light Scattering Spectroscopy of Propagating Magnons at Sub-Kelvin Temperatures}

\author{David Schmoll\href{https://orcid.org/0000-0001-5260-2052}{\includegraphics[scale=0.02]{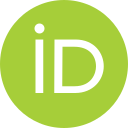}}}
\email{david.schmoll@univie.ac.at}
\affiliation{Faculty of Physics, University of Vienna, Boltzmanngasse 5, 1090, Vienna, Austria}
\affiliation{Vienna Doctoral School in Physics, University of Vienna, Boltzmanngasse 5, 1090, Vienna, Austria}
\author{Nikolai Kuznetsov\href{https://orcid.org/0000-0001-8650-1954}{\includegraphics[scale=0.02]{orcid.png}}}
\affiliation{NanoSpin, Department of Applied Physics, Aalto University School of Science, P.O. Box 15100, 00076 Aalto, Finland}
\author{Phillip Rehberger}
\affiliation{Faculty of Physics, University of Vienna, Boltzmanngasse 5, 1090, Vienna, Austria}
\author{Franz Vilsmeier\href{https://orcid.org/0000-0002-5648-3051}{\includegraphics[scale=0.02]{orcid.png}}}%
\affiliation{Faculty of Physics, University of Vienna, Boltzmanngasse 5, 1090, Vienna, Austria}
\author{Roman Verba\href{https://orcid.org/0000-0001-8811-6232}{\includegraphics[scale=0.02]{orcid.png}}}%
\affiliation{V.G. Baryakhtar Institute of Magnetism of the NAS of Ukraine, 03142, Kyiv, Ukraine}
\author{Denys Slobodianiuk\href{https://orcid.org/0000-0001-8908-4703}{\includegraphics[scale=0.02]{orcid.png}}}%
\affiliation{V.G. Baryakhtar Institute of Magnetism of the NAS of Ukraine, 03142, Kyiv, Ukraine}
\author{Rostyslav O. Serha\href{https://orcid.org/0009-0007-2962-1109}{\includegraphics[scale=0.02]{orcid.png}}}%
\affiliation{Faculty of Physics, University of Vienna, Boltzmanngasse 5, 1090, Vienna, Austria}
\author{Khrystyna O. Levchenko\href{https://orcid.org/0000-0002-0903-5942}{\includegraphics[scale=0.02]{orcid.png}}}%
\affiliation{Faculty of Physics, University of Vienna, Boltzmanngasse 5, 1090, Vienna, Austria}
\author{Sebastiaan van Dijken\href{https://orcid.org/0000-0001-6372-2252}{\includegraphics[scale=0.02]{orcid.png}}}%
\affiliation{NanoSpin, Department of Applied Physics, Aalto University School of Science, P.O. Box 15100, 00076 Aalto, Finland}
\author{Andrii V. Chumak\href{https://orcid.org/0000-0001-5515-0848}{\includegraphics[scale=0.02]{orcid.png}}}%
\email{andrii.chumak@univie.ac.at}
\affiliation{Faculty of Physics, University of Vienna, Boltzmanngasse 5, 1090, Vienna, Austria}
\author{Sebastian Knauer\href{https://orcid.org/0000-0002-5790-4575}{\includegraphics[scale=0.02]{orcid.png}}}%
\email{knauer.seb@gmail.com}
\affiliation{Faculty of Physics, University of Vienna, Boltzmanngasse 5, 1090, Vienna, Austria}
\affiliation{Center for Digital Safety \& Security - Optical Quantum Technologies, AIT Austrian Institute of Technology GmbH, Giefinggasse 4, 1210, Vienna, Austria}

\begin{abstract}

Coupling light to magnetic excitations in the form of spin waves underpins both the optical study of magnetism and emerging schemes for quantum transduction, positioning the quanta of these excitations, magnons, as promising carriers for hybrid quantum networks. However, exploiting them in the quantum regime requires millikelvin temperatures to suppress thermal magnon populations, thereby confining such experiments to dilution refrigerators. There, magnons can already be excited and read out electrically, yet an optical interface required for microwave-to-optical photon conversion has been missing. Here, we demonstrate the first optical detection of coherently driven, propagating spin waves via Brillouin Light Scattering (BLS) spectroscopy inside a dilution refrigerator. By simultaneously recording the optical and electrical responses of the same spin-wave mode in a yttrium iron garnet film, we find that the BLS spectra track the electrically measured transmission across a range of applied magnetic fields. For the lowest optical power of $(7.9\pm 1)$~\qty{}{\micro\watt} that still enabled spin-wave detection, we measured a global equilibrium sample temperature of \qty{510}{\milli\kelvin} via a resistance thermometer, while numerical modelling of the laser-induced heating yields a maximum local temperature of \qty{900}{\milli\kelvin} at the focal spot. This brings free-space optical access to magnons into the sub-kelvin regime, representing a milestone towards magnon-mediated quantum transduction in hybrid quantum systems.

\end{abstract}

\maketitle

\section{Introduction}

Magnonics deals with spin waves and their quanta, magnons, as the eigenexcitations of the collective spins of magnetically ordered media~\cite{Pirro2021, Serga2010}. Their rich dispersion characteristics, together with continuous advances in nanofabrication and the rapidly increasing demand for innovation in data transportation and processing, have sparked interest in the potential of spin waves as data carriers in novel computing schemes~\cite{Zeenba2024, Finocchio2024, Wang2024, Chumak2022, Mahmoud2020}. Practical on-chip applications require the ability to actively control, excite, and detect spin waves with high spatial and temporal resolution. In this regard, optical techniques such as Brillouin Light Scattering (BLS) spectroscopy have enabled numerous pioneering experiments~\cite{Kuznetsov2025, Krcma2025, Heinz2020, Vogel2018}. BLS measures the frequency shift of light after inelastic scattering from spin waves or phonons and is a well-established experimental technique in the field of magnonics~\cite{Kargar2021, Sebastian2015, Demokritov2008, Sandercock1973}, but also for interdisciplinary applications beyond physics~\cite{Falkner2025_BrillouinCancer, Bouvet2025, Antonacci2020, Bevilacqua2023_LineScanBrillouin, Keshmiri2024_BrillouinAnisotropy}.

Beyond the classical encoding of information in the phase and amplitude of spin waves, the quantum nature of magnons has recently moved to the forefront of the field. Their broad frequency range (GHz--THz) and a set of experimentally accessible parameters that enable manipulation of the dispersion relation offer great potential for coherent interactions between magnons and a variety of other physical platforms, such as photons, phonons, and superconducting qubits~\cite{Serha2026, Serha2025_YSGAG, Jiang2023, Li2020, Forsch2020, Quirion2019}. These properties position magnons as a strong contender for quantum transduction in scalable hybrid quantum networks, offering promising opportunities for quantum telecommunication~\cite{Wehner2018_QuantumInternet, Kimble2008_QuantumInternet, Reiserer2015}. Inspired by the successful coupling of magnons to superconducting qubits~\cite{Nakamura2020, Tabuchi2015}, achieving optical access to magnons at millikelvin temperatures would unlock an additional coupling mechanism and pave the way for future hybrid opto-magnonic experiments~\cite{BHOI_book, BHOI_book2} that fully utilise propagating magnons at the quantum level with spatially separated sources and detectors~\cite{Knauer_2023, vanLoo2018}.

A key challenge in quantum magnonics remains in extending magnon lifetimes to match those of superconducting circuits, for which long coherence times are vital in quantum information applications~\cite{Krantz2019}. This mandates materials with minimised magnetic damping and the suppression of thermal magnon populations, requiring temperatures in the millikelvin regime, according to the Bose-Einstein occupation statistics~\cite{Serha2026_APL, Serha2026}. Experiments in such low-temperature environments are performed in Dilution Refrigerator (DR) systems, which significantly complicates the operation of established room-temperature techniques. While electrical methods for spin-wave excitation and detection have been successfully transformed to millikelvin conditions~\cite{Schmoll2025_Wavenumber, Serha2025_DampingEnhancement}, optical detection techniques such as BLS spectroscopy have been limited to liquid-helium temperatures~\cite{Demokritov1986, Demokritov1987, Che2025} and were not yet implemented in dilution refrigerators. 

Here, we overcome this long-standing limitation and report the first Brillouin light scattering spectroscopy of propagating magnons inside a dilution refrigerator, extending free-space optical access to the sub-kelvin regime. Spin waves were electrically excited by placing a Yttrium Iron Garnet (YIG) film on two microwave antennas. Simultaneously, a \qty{532}{\nano\meter} Continuous-Wave~(CW) laser was focused on the sample while the inelastically backscattered light was analysed using a Tandem Fabry Pérot Interferometer~(TFPI). This experimental configuration yielded a direct overlap of the electrical and optical transmission signals, unambiguously confirming the detection of the same coherently driven spin-wave modes. At the lowest optical powers, we measured a global equilibrium sample temperature of $T_\mathrm{p}^\mathrm{global}~=$~\qty{510}{\milli\kelvin} via a resistance thermometer. The experiments were combined with numerical simulations of the laser-induced heating at the focal spot with the software packages \textit{Ansys Lumerical} and \textit{COMSOL Multiphysics}, demonstrating that the local temperature $T_\mathrm{p}^\mathrm{local}~\approx$~\qty{900}{\milli\kelvin} remains in the sub-kelvin regime.

\section{Methods}
\label{sec:modeling}
\subsection{\label{subsec:micromag}Experimental Implementation}

Spin waves were excited and detected in a one-sided, \qty{18}{\micro\meter}-thick YIG film, grown on a \qty{500}{\micro\meter}-thick Gadolinium Gallium Garnet (GGG) substrate. To enable simultaneous optical and electrical detection, we coherently excited spin waves in the Magnetostatic Surface Spin Wave (MSSW) configuration ($B_0 \perp k$) using a gold-plated copper microstrip antenna patterned on an Aluminium Nitride (AlN) substrate electrically connected to a Vector Network Analyser (VNA) via RF transmission lines integrated into the dilution refrigerator. The underside of the AlN substrate is copper-plated and glued to an oxygen-free copper adapter using silver paint. The adapter itself is thermally anchored to the dilution refrigerator's sample holder puck and cooled by the mixing chamber. A ruthenium oxide resistance thermometer, positioned on the sample puck close to the copper adapter, measures the global equilibrium sample temperature $T_\mathrm{p}^\mathrm{global}$. The YIG film is in contact with the AlN substrate and affixed at the edges using GE varnish infused with AlN powder to reduce interfacial thermal resistance. A second microstrip antenna, separated by \qty{2}{\milli\meter} from the excitation antenna, enables the electrical detection of the propagating magnons. For simultaneous optical detection, the laser light was focused through the GGG substrate onto the YIG film, with the focal spot positioned between the two antennas.
\begin{figure}[t]
	\includegraphics[width=0.45\textwidth]{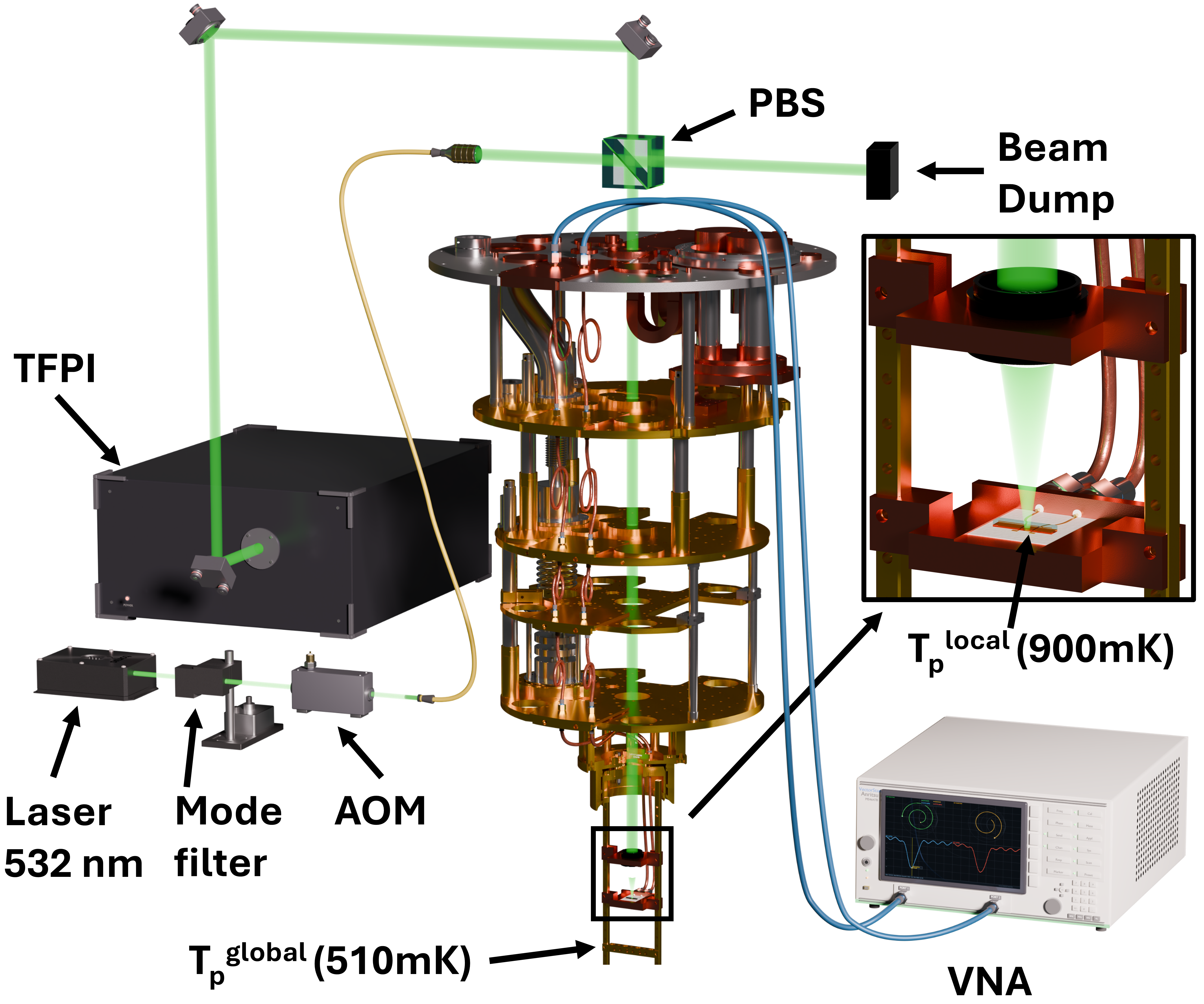}
    \captionsetup{justification=justified}
	\caption{\justifying Schematic illustration of the implementation of a dilution refrigerator system in a backscattering BLS spectroscopy assembly.~$T_\mathrm{p}^\mathrm{global}$ and $T_\mathrm{p}^\mathrm{local}$ indicate the global equilibrium temperature measured with a resistance thermometer and the numerically simulated local temperature at the laser focal spot for the lowest applied optical power.} 
	\label{fig:SetupIllustration}
\end{figure}

The optical detection of spin waves at dilution refrigerator temperatures requires integrating the cryostat into a conventional BLS spectroscopy setup. The DR-system itself can reach a base temperature of \qty{10}{\milli\kelvin} and contains a superconducting vector magnet. The cryogenic environment is maintained under vacuum (\qty{e-6}{\milli\bar}) and organised into different thermal stages, with optical access ports in the centre. The presented BLS assembly comprises a CW laser source of wavelength $\lambda =$ \qty{532}{\nano\meter}, a laser mode filter, a tandem Fabry Pérot interferometer, and a Single Photon Avalanche Detector (SPAD), all placed at room temperature on an actively vibration-isolated optical table. To control the heat load introduced into the dilution refrigerator, the initial laser beam can be attenuated using a neutral-density filter with variable optical density and additionally can be pulsed by an Acousto-Optic Modulator (AOM) with a minimum rise time of \qty{25}{\nano\second}. After the AOM, the laser beam is coupled into a single-mode optical fibre that connects the optical table and the top plate of the dilution refrigerator. On top of the cryostat, the light is converted back to a free beam and guided into the cryogenic environment through an anti-reflection coated \qty{4}{\milli\meter}-thick UV-fused Silica (UVFS) window via a polarising beam splitter (PBS) cube. Two \qty{1}{\milli\meter}-thick anti-reflection coated NBK7 windows at the \qty{50}{\kelvin} and \qty{4}{\kelvin} stage suppress thermal black-body radiation inside the dilution refrigerator and allow the system to reach base temperature despite the optical window on top. The incident laser beam is focused onto the investigated YIG film through the GGG substrate via a 1-inch plano-convex UVFS-lens with a focal length of $f=$ \qty{35}{\milli\meter} and a Numerical Aperture (NA) of $\approx~0.34$. The backscattered light within the NA of the lens is coupled back out of the cryostat, guided to the optical table as a free beam, and collected at the TFPI, enabling the detection of the induced frequency shift due to the inelastic scattering on spin waves or phonons~\cite{Sandercock1973}. An illustration of the essential components of the dilution refrigerator coupled BLS is depicted in Fig.~\ref{fig:SetupIllustration}.
\begin{figure*}[t]
	\includegraphics[width=1\textwidth]{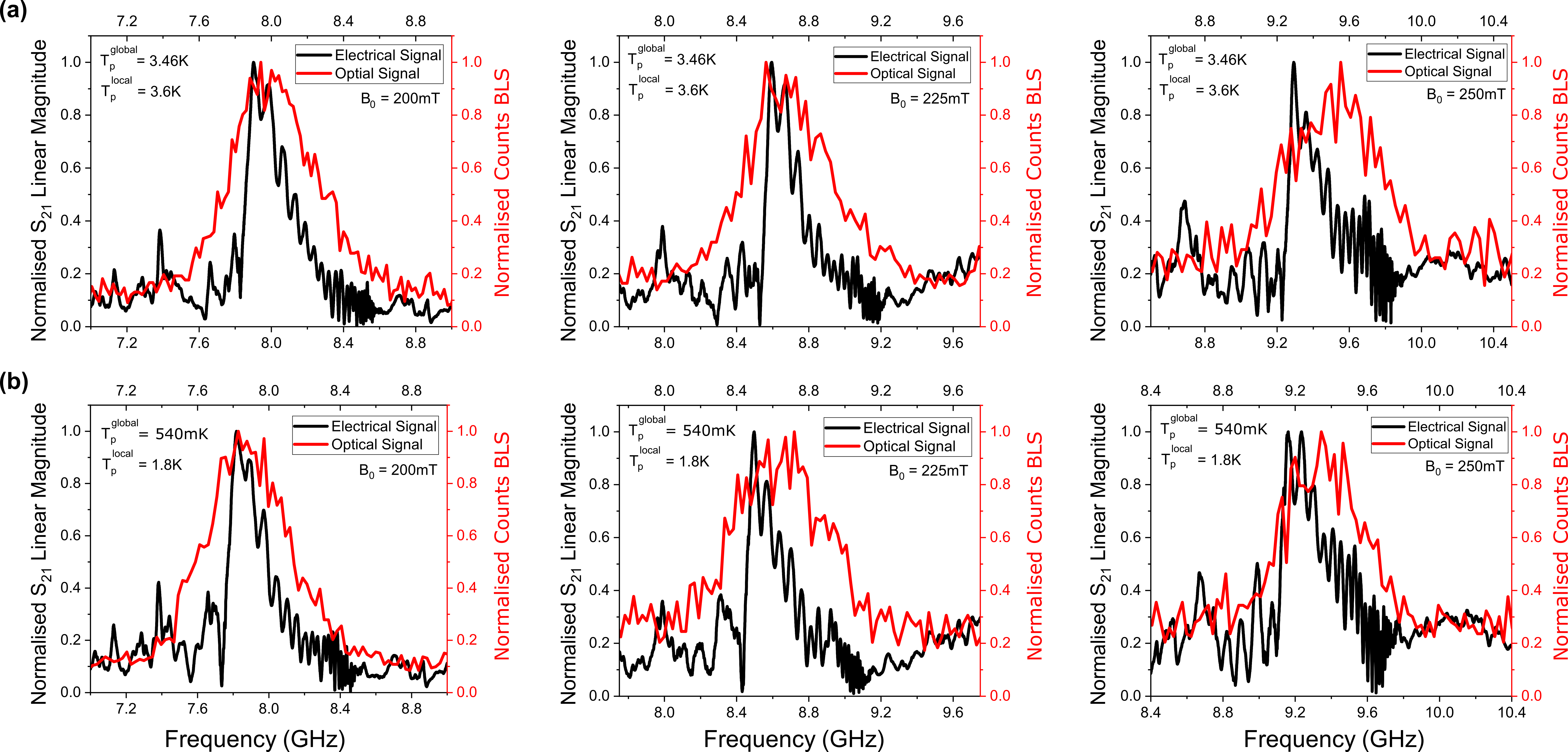}
    \captionsetup{justification=justified}
	\caption{\justifying Normalised spin-wave transmission spectra recorded electrically with Propagating Spin-Wave Spectroscopy (PSWS, black curves) and optically via Brillouin Light Scattering spectroscopy (BLS, red curves). Measurements were performed with an incident laser power at the YIG film of $(79\pm2)$~\qty{}{\micro\W} for three different magnetic fields and \textbf{(a)} pulse tube operation only and \textbf{(b)} dilution cooling. To provide an absolute reference for the normalised data at a magnetic field of \qty{200}{\milli\tesla} and $T_\mathrm{p}^\mathrm{global}~=$~\qty{540}{\milli\kelvin}, the maximum measured spin-wave amplitude reached \qty{-8.2}{\decibel} for the electric measurement and 212 counts for the optically acquired BLS spectrum.} 
	\label{fig:ComparisonSpectra100uW}
\end{figure*}

\subsection{Numerical Simulation of the Local Temperature Gradient}
\label{sec:numsim}

Laser-induced local heating during DR-coupled BLS spectroscopy was modelled using integrated three-dimensional (3D) optical and thermal simulations. The spatial distribution of absorbed optical power was first computed using the Finite-Difference Time-Domain (FDTD) solver in the \textit{Ansys Lumerical} software package. The simulated geometry consisted of an \qty{18}{\micro\m}-thick YIG film on a GGG substrate. A Gaussian beam with a wavelength of \qty{532}{\nano\m} and a waist diameter of $\approx$~\qty{5}{\micro\m} was focused through the GGG substrate onto the YIG layer, consistent with the back-scattering BLS geometry. The volumetric absorbed power density in the YIG film was extracted using an internal \textit{Lumerical} monitor and exported as a 3D heat-source profile for subsequent thermal analysis.

The 3D heat-source profile was imported into \textit{COMSOL Multiphysics} and applied to the full GGG/YIG/AlN/Cu multilayer stack using the heat transfer in solids module. Thermal material parameters were taken from Refs.~\cite{Boona2014_PRB, bhandari1966scattering, walton1973thermal, pan2013specific, daudin1982thermodynamic, prakash2018evidence, simon1994_AlN} and the temperature-dependent values adopted for YIG and GGG are displayed in Fig.~S2 of the Supplementary Information~\cite{supplement}.

Steady-state simulations were performed to determine the equilibrium temperature distribution $T_\mathrm{p}^\mathrm{local}$ at the laser focal spot under continuous-wave illumination. A fixed-temperature boundary condition, set equal to the experimentally measured global equilibrium temperature $T_\mathrm{p}^\mathrm{global}$ of the ruthenium oxide resistance thermometer, was imposed at the bottom Cu surface. All remaining external boundaries were treated as adiabatic. The resulting temperature distribution was analysed in the $xz$-plane passing through the centre of the heat source.

To characterise the heating dynamics under pulsed illumination, the transient solver was employed. The fixed-temperature boundary condition was replaced by a heat-flux boundary condition that accounts for the finite cooling power of the dilution refrigerator: \[q = h(T_\mathrm{p}^\mathrm{global} - T),\] where $h$ is an effective heat-transfer coefficient and $T_\mathrm{p}^\mathrm{global}$ is the experimentally measured global temperature of the thermometer. The heat source was activated for \qty{1}{\micro\s} with a pulse period of \qty{4}{\micro\s}, and the temporal evolution of the maximum temperature within the simulated structure was recorded. Full details of both the optical and thermal simulation procedures, along with an extensive description of the experimental setup, are provided in the Supplementary Information~\cite{supplement}.

\section{Results}

\subsection{Comparison of Optical and Electrical Transmission Spectra}
\begin{figure*}[t]
	\includegraphics[width=0.92\textwidth]{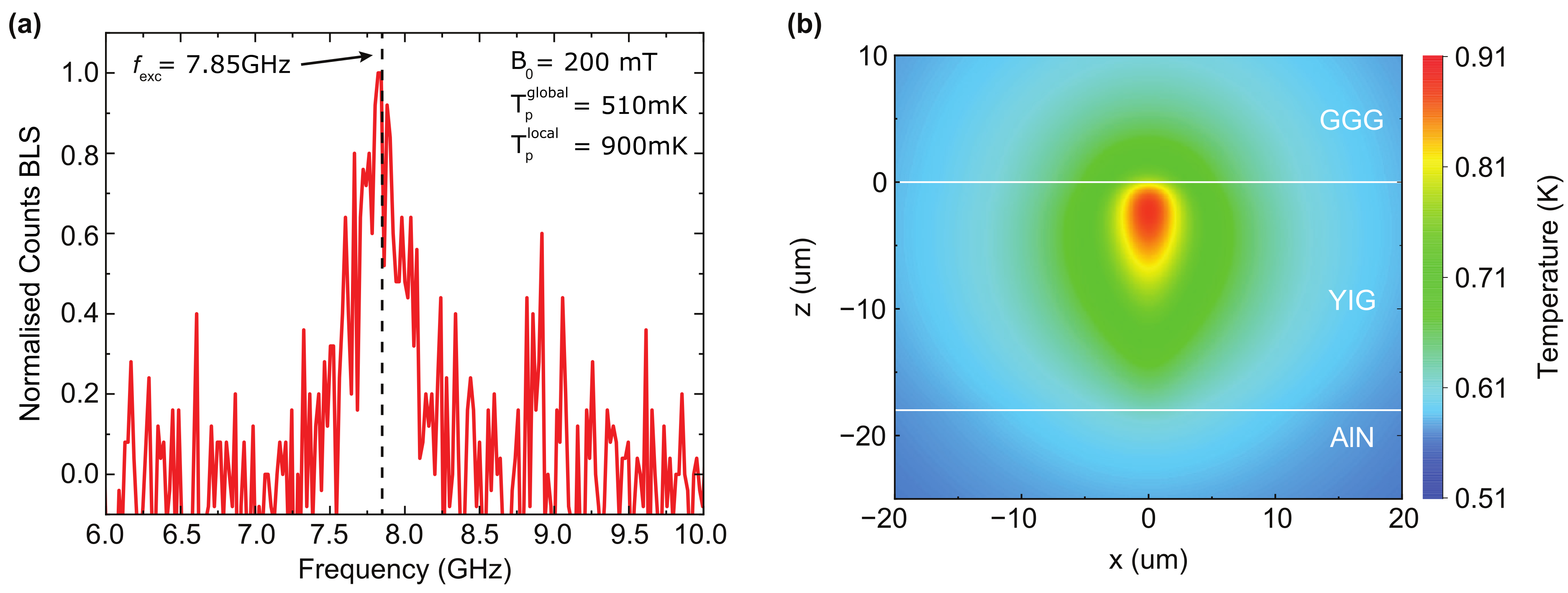}
    \captionsetup{justification=justified}
	\caption{\justifying \textcolor{black}{\textbf{(a)} Normalised BLS intensity spectrum recorded for continuous illumination at a fixed electric excitation frequency of \qty{7.85}{\giga\hertz} (dashed black line) with a bias field of \qty{200}{\milli\tesla}. The laser power incident on the YIG film was $(7.9\pm 1)$~\qty{}{\micro\watt}, yielding a global equilibrium temperature $T_\mathrm{p}^\mathrm{global}$~=~\qty{510}{\milli\kelvin} at the thermometer and a local temperature $T_\mathrm{p}^\mathrm{local}~\approx$~\qty{900}{\milli\kelvin} at the focal spot, according to the numerical model. \textbf{(b)} Simulated steady-state temperature distribution in the $xz$-plane in the vicinity of the laser focal spot in the GGG/YIG/AlN stack under CW illumination at the given experimental parameters.}}
	\label{fig:3}
\end{figure*}

Figure~\ref{fig:ComparisonSpectra100uW} displays the normalised spin-wave transmission spectra obtained with Propagating Spin-Wave Spectroscopy (PSWS, black curves) and via Brillouin Light Scattering spectroscopy (BLS, red curve) for the external magnetic fields $B_0$ =~\qty{200}{\milli\tesla},~\qty{225}{\milli\tesla}, and \qty{250}{\milli\tesla} in the MSSW configuration. The one-to-one correspondence between the electrical and optical spectra across all probed resonance conditions, set by the external magnetic field, unambiguously identifies the backscattered light with the coherently driven magnon mode and constitutes the first detection of propagating spin waves by BLS spectroscopy in a dilution refrigerator. As expected from the governing dispersion relations, the maximum spin-wave transmission shifts to higher frequencies with increasing magnetic field, moving from approximately $\qty{7.9}{\giga\hertz}$ at $\qty{200}{\milli\tesla}$ to $\qty{9.6}{\giga\hertz}$ at $\qty{250}{\milli\tesla}$. During the BLS acquisition, spin waves were continuously excited with a microstrip antenna by sweeping the applied microwave frequency over a \qty{500}{\mega\hertz}-wide band around the transmission maximum with \qty{-5}{\dBm} microwave power at the antenna and an Intermediate Frequency Bandwidth (IFBW) of \qty{1}{\kilo\hertz}. To optically detect the coherently excited spin-wave modes, a laser was continuously focused onto the film approximately \qty{1}{\milli\meter} from the excitation antenna with $(79~\pm~2)$~\qty{}{\micro\W} incident optical power at the YIG surface. The electric signature of the spin wave was recorded via the $S_{21}$-parameter of the VNA for a frequency band of \qty{5}{\giga\hertz} spanning around the transmission maximum. 

To demonstrate the functionality of the experimental platform, the measurements were performed for two distinct cooling regimes, in the pulse-tube-only operation (Fig.~\ref{fig:ComparisonSpectra100uW}~(a)) and the operation of the dilution cycle (Fig.~\ref{fig:ComparisonSpectra100uW}~(b)). For a continuous feed of microwave and laser power, the equilibrium global temperature $T_\mathrm{p}^\mathrm{global}$ measured with the ruthenium oxide resistance thermometer stabilises at \qty{3.46}{\kelvin} for the pulse-tube-only case and at \qty{540}{\milli\kelvin} for the dilution cooling case. As the temperature at the thermometer does not necessarily resemble the local temperature gradient directly at the focal spot of the laser, we utilised numerical simulations to estimate the temperature in the optical probing area, yielding an increased local temperature $T_\mathrm{p}^\mathrm{local}$ of \qty{3.6}{\kelvin} for the pulse-tube-only case and \qty{1.8}{\kelvin} for the dilution cooling. Graphical illustrations of the local temperature distributions, extracted from the numerical model, are provided in Fig.~S3 in the Supplementary Information~\cite{supplement}.

\subsection{Steady-State BLS Spectroscopy of Magnons with Minimal Optical Power}

To minimise local laser heating, optical measurements were conducted with an incident laser power of $(7.9~\pm~1)$~\qty{}{\micro\watt} at the YIG film. Simultaneously, spin waves were driven at a fixed frequency of \qty{7.85}{\giga\hertz} with a microwave power of \qty{-5}{\dBm} at the antenna. To improve excitation efficiency, the VNA was operated in CW mode rather than using a frequency sweep. An external magnetic field of magnitude $B_0 =$~\qty{200}{\milli\tesla} was applied in the MSSW configuration. Figure ~\ref{fig:3} (a) depicts the normalised optical spin-wave spectrum acquired by the TFPI, indicating a clear maximum of BLS counts at the electrical excitation frequency. The global temperature recorded by the thermometer $T_\mathrm{p}^\mathrm{global}$ for these experimental conditions was \qty{510}{\milli\kelvin}, while the numerical simulations of the local temperature gradient $T_\mathrm{p}^\mathrm{local}$, illustrated in Fig. ~\ref{fig:3} (b), yield $\approx$~\qty{900}{\milli\kelvin}. Hence, even with a deliberately simple collection optic, the dilution-refrigerator-coupled BLS already reaches the sub-kelvin regime.

\subsection{Pulsed BLS Spectroscopy of Magnons}
\begin{figure*}[t]
	\includegraphics[width=0.92\textwidth]{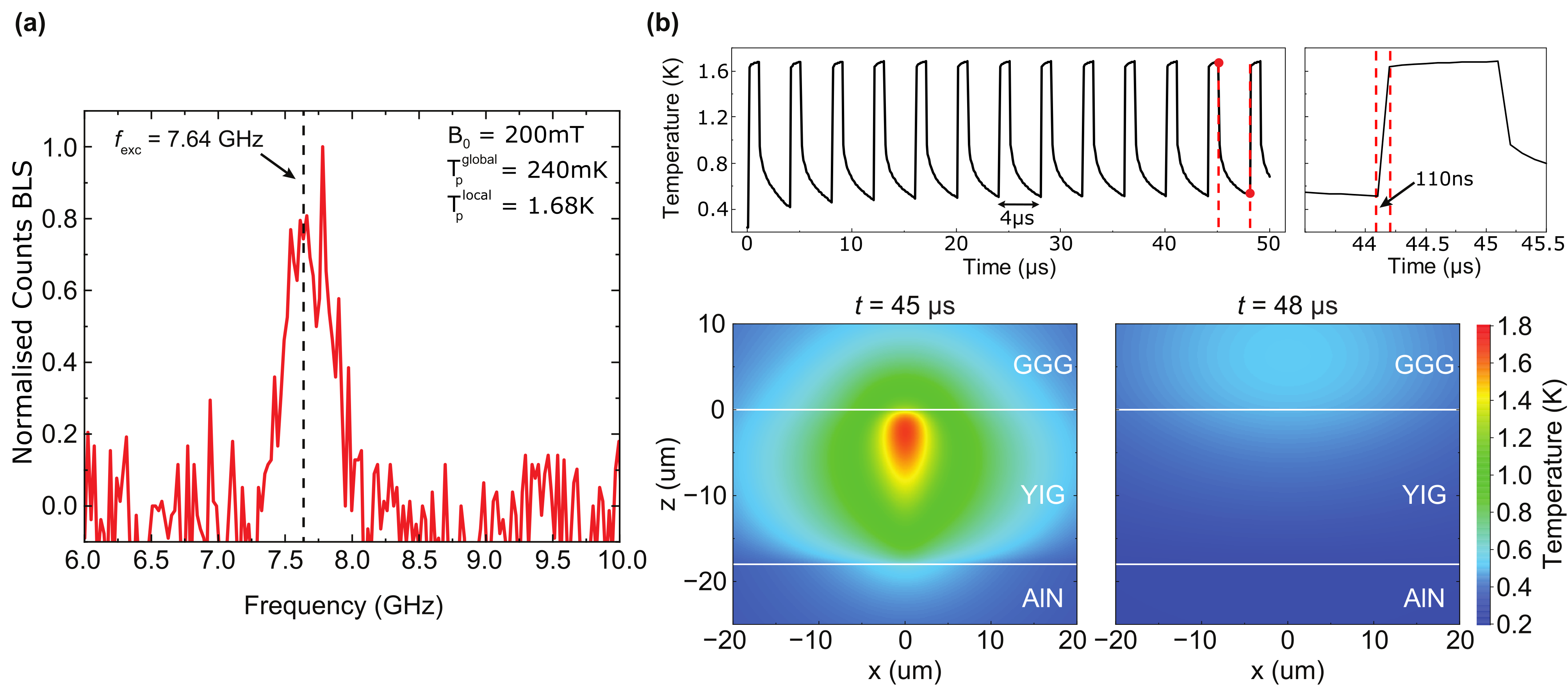}
    \captionsetup{justification=justified}
	\caption{\justifying \textcolor{black}{\textbf{(a)} Normalised BLS intensity spectrum recorded in the pulsed-laser regime at a fixed electric excitation frequency of \qty{7.64}{\giga\hertz} (black dashed line) with a bias field of \qty{200}{\milli\tesla}. The average laser power incident at the YIG film was \qty{20}{\micro\watt}, yielding an equilibrium temperature of $T_\mathrm{p}^\mathrm{global}$~=~\qty{240}{\milli\kelvin} at the thermometer and $T_\mathrm{p}^\mathrm{local}$~=~\qty{1.68}{\kelvin} at the focal spot, according to the numerical model. \textbf{(b)} Transient thermal simulations of laser-pulse-induced local heating under the given experimental conditions. Top: temporal evolution of the peak temperature in the GGG/YIG/AlN multilayer stack during periodic \qty{1}{\micro\s} laser pulses with a \qty{4}{\micro\s} period and the zoom into one pulse. The red dots indicate the instants $t=45$~\qty{}{\micro\s} and $t=48$~\qty{}{\micro\s} selected for the spatial maps. Bottom: simulated temperature distributions in the $xz$-plane in the vicinity of the laser focal spot for the laser-ON ($t=45$~\qty{}{\micro\s}) and laser-OFF ($t=48$~\qty{}{\micro\s}) states.}}
	\label{fig:4}
\end{figure*}

To reduce the average heat load on the DR compared with the continuous feed of laser power and electric excitation, measurements were also performed in the pulsed regime by modulating the laser signal using an AOM. A microwave power of \qty{-10}{\dBm} was applied by a signal generator instead of a VNA, enabling signal modulation via an external TTL trigger synchronised with the AOM transducer. The pulse duration was set to \qty{1}{\micro\second} with a period of \qty{4}{\micro\second}, yielding a repetition rate of \qty{250}{\kilo\hertz}. Spin waves were electrically excited at a fixed frequency of \qty{7.64}{\giga\hertz} in the MSSW configuration with an externally applied magnetic field of magnitude $B_0 =$~\qty{200}{\milli\tesla}. Figure ~\ref{fig:4} (a) displays the normalised, optically acquired spin-wave spectrum for an average laser power of \qty{20}{\micro\watt} incident at the YIG film, exhibiting a pronounced peak in BLS counts at the excitation frequency. Under these pulsed experimental conditions, the global temperature recorded by the thermometer $T_\mathrm{p}^\mathrm{global}$ was \qty{240}{\milli\kelvin}, whereas numerical simulations yield a maximum local temperature $T_\mathrm{p}^\mathrm{local}$ of \qty{1.68}{\kelvin} at the laser focal spot. Figure ~\ref{fig:4} (b) shows the simulated temporal evolution of the maximum temperature within the multilayer stack. During the laser-ON interval, the local temperature abruptly rises to \qty{1.68}{\kelvin} in a matter of only $\approx$~\qty{110}{\nano\second}, as illustrated in the zoomed-in temporal evolution of $T_\mathrm{p}^\mathrm{local}$. Upon laser extinction, it decays to approximately \qty{510}{\milli\kelvin} during the laser-OFF interval. The spatial temperature distributions computed at $t~=~45$~\qty{}{\micro\s} and $t~=~48$~\qty{}{\micro\s} confirm that the strongly localised heating generated during illumination diffuses predominantly into the GGG substrate once the laser is switched OFF, consistent with the low thermal conductivity of the YIG film relative to the substrate at cryogenic temperatures.

\section{Discussion}
\label{sec:conclusion}

In summary, we have established Brillouin light scattering spectroscopy as a direct, free-space optical probe of coherently driven propagating magnons inside a dilution refrigerator, transferring a cornerstone technique of magnonics into the sub-kelvin regime for the first time. We electrically excited spin waves via a microstrip antenna in a YIG film grown on GGG, and observed excellent agreement between conventional electrical spin-wave transmission spectra and optically acquired BLS counts across multiple magnetic fields, unambiguously identifying the backscattered light with the propagating magnon mode. Because the same assembly can detect phonons simply by replacing the polarising beam splitter with a conventional one, it further opens access to magnomechanical coupling and phonon spectroscopy within the same cryogenic environment.

While BLS spectroscopy of magnons was realised at liquid-helium temperatures~\cite{Demokritov1986, Demokritov1987}, its implementation under dilution-refrigerator conditions, with simultaneous optical and electrical readout, has, to our knowledge, not been demonstrated before. Decreasing the incident laser power at the YIG film to $(7.9\pm 1)$~\qty{}{\micro\watt} yields a temperature of $T_\mathrm{p}^\mathrm{local}~\approx$~\qty{900}{\milli\kelvin} at the laser focal spot according to numerical simulations, suggesting that the DR-coupled BLS setup already enables access to the sub-kelvin regime. 

By switching from a continuous to a pulsed optical power feed via an AOM, the average heat load on the dilution refrigerator is reduced. However, numerical simulations of the temporal temperature evolution at the laser focal spot reveal that $T_\mathrm{p}^\mathrm{local}$ remains dominated by the peak power of the incident optical pulses, owing to the rapid drop $C\propto T^3$ of the heat capacity in YIG with temperature according to the Debye model. Overcoming this localised heating effect would require non-insulating materials with higher heat capacities than YIG, or shorter pulse lengths below \qty{100}{\nano\second}, which exceed the practical capabilities of the AOM used in this study.

Furthermore, the numerical modelling highlights the localised hotspot induced by laser heating. Because the GGG substrate is transparent at the given wavelength, the incident light is almost entirely absorbed within the YIG layer. Similar to the heat capacity, the thermal conductivity of YIG decreases as $T^3$ well below the Debye temperature. According to the numerical model, this severe insulating effect creates a steep spatial temperature gradient on the order of tens of micrometres, in contrast to room-temperature experiments, where heat readily diffuses along the film interface~\cite{Vogel2015}. For the low-temperature case, the numerical model indicates that the localised heat primarily dissipates into the GGG substrate due to its comparatively higher heat capacity and thermal conductivity. Temperature-dependent values for both the heat capacity and the thermal conductivity in YIG and GGG are provided in Fig.~S2 in the Supplementary Information~\cite{supplement}.

Since the central motivation for cryogenic magnonics is access to the quantum regime, it is instructive to relate the temperatures reached here to the single-magnon limit. Following Bose-Einstein statistics, the thermal occupation of an individual magnon level near the ferromagnetic resonance falls below one for a temperature of approximately \qty{530}{\milli\kelvin} and therefore close to the local temperature $T_\mathrm{p}^\mathrm{local}~\approx$~\qty{900}{\milli\kelvin} attained in the steady-state measurements. To quantify the thermal background concretely, we estimate the thermal magnon density contained in the optically probed volume $V\approx3.5\times10^{-10}$~\qty{}{\centi\meter^3}, set by the focal spot with a diameter of $\approx$~\qty{5}{\micro\meter} across the \qty{18}{\micro\meter} film thickness. At $T_\mathrm{p}^\mathrm{local}~\approx$~\qty{900}{\milli\kelvin}, this volume holds of the order of $1\times10^7$ thermal magnons in total. Of this population, only $\approx5\times10^4$ fall within the BLS detection band, and merely $\approx4\times10^2$ lie within the much narrower ferromagnetic-resonance linewidth that is relevant for coherent operation. This pronounced reduction shows how strongly the finite detection bandwidth concentrates the measurement around the coherent mode. These numbers demonstrate that reaching the single-magnon regime, in which the magnon number integrated over the probed volume and bandwidth drops below one, is more demanding and would require temperatures below \qty{28}{\milli\kelvin} for the YIG film used. In addition, the presented experiments probed coherently driven spin-wave modes for which the magnon density significantly exceeds the thermal population, independent of temperature.  

However, the single-magnon requirement is not a fundamental barrier since the magnon number scales as $N_\mathrm{th} = n_\mathrm{th}V$ with the mode volume $V$. Hence, confining the magnon population in a typical nanoscale waveguide of size \qty{100}{\nano\meter}$\times$\qty{1}{\micro\meter}$\times$\qty{100}{\micro\meter} already shifts the temperature to satisfy $N_\mathrm{th} < 1$ at $\approx$~\qty{200}{\milli\kelvin}. Besides reducing the sample volume, another effective strategy is to operate at higher magnon frequencies. Since increasing the FMR frequency $\omega_0$ by a factor $p$ raises all characteristic temperatures by the same factor, working with \qty{50}{\giga\hertz} magnons, for example, would shift the single-magnon BLS condition to $\approx$~\qty{0.3}{\kelvin} and to $\approx$~\qty{0.6}{\kelvin} when combined with a reduced film thickness. Minimisation of the absorbed laser power in the YIG layer, either by shifting to longer wavelengths or by decreasing the film thickness, would reduce local heating and enable optical magnon detection at such temperatures. These strategies underscore that the single-magnon regime is within practical reach for the presented experimental platform. A detailed analysis of the thermal and coherently excited magnon populations is provided in the Supplementary Information and illustrated in Fig.~S4~\cite{supplement}.

The results presented were obtained with a deliberately minimal optical assembly, employing a simple lens to focus and collect the backscattered light. While a lens with a longer focal length and a smaller NA would spread the optical power density and reduce local heating, a cryogenic objective, combined with a nanopositioning stage, would increase spatial resolution and improve signal collection. Coupling this advanced optical hardware with highly efficient magneto-optical materials, such as bismuth-substituted YIG, and optimised geometries, such as optical cavities, will significantly boost overall experimental performance. Such improvements in sensitivity, together with a decrease of the magnon population below $N < 1$, will establish a robust platform for future hybrid opto-magnonic experiments, coherent magnon-photon transduction, and the potential utilisation of magnons in quantum telecommunication networks.
\vspace{\baselineskip}

\section*{Acknowledgements}
\label{sec:acknowledgements}

This research was funded in whole or in part by the Austrian Science Fund (FWF) project No. 10.55776/I6568. R.V. and D.Sl. were supported by the NAS of Ukraine, project No. 0123U104827. K.O.L acknowledges the Austrian Science Fund (FWF) project No. 10.55776/ESP526. This research received funding from the European Union's Horizon 2020 research and innovation programme under the Marie Sklodowska–Curie grant agreement No. 101025758, project OMNI. The numerical simulations presented in this research were performed using computational resources provided by the Aalto Science-IT project.~S.K. thanks Silvia Viola Kusminskiy and Sanchar Sharma for useful discussions. 

\section*{Contribution}

D.S.~planned the experiment, conducted all measurements, processed and analysed the experimental data, and authored the initial draft of the manuscript in consultation with S.K. and N.K.; N.K.~established the numerical model and analysed the simulations of the local temperature distribution. P.R.~and F.V.~supported the construction of the experimental setup. R.V. and D.Sl. formulated the theory and analysis of the estimation of magnon populations. R.O.S.~assisted in interpreting the experimental results and contributed to the graphical illustrations. K.O.L. supported the room temperature characterisation of the sample. S.v.D. supervised the numerical simulations. A.V.C. supervised the measurements, provided the experimental infrastructure, and contributed to the analysis.~S.K. had the initial idea for the millikelvin BLS, supervised the measurements, and led the project. All authors discussed the results and contributed to the manuscript. The authors declare no competing interests.

\vspace{2cm}



\bibliographystyle{apsrev4-2}
\bibliography{bibliography}

@article{Pirro2021,
    author = {Philipp Pirro and Vitaliy I. Vasyuchka and Alexander A. Serga and Burkard Hillebrands},
    journal = {Nat. Rev. Mater.},
    title = {Advances in coherent magnonics},
    volume = {6},
    pages = {1114-1135},
    year = {2021},
    doi = {10.1038/s41578-021-00332-w},
}

@article{Serga2010,
    author = {A. A. Serga and A. V. Chumak and B. Hillebrands},
    journal = {J. Phys. D: Appl. Phys.},
    title = {{YIG} magnonics},
    volume = {43},
    eid = {264002},
    year = {2010},
    doi = {10.1088/0022-3727/43/26/264002},
}

@article{Zeenba2024,
    author = {Zenbaa, Noura and Abert, Claas and Majcen, Fabian and Kerber, Michael and Serha, Rostyslav O. and Knauer, Sebastian and Wang, Qi and Schrefl, Thomas and Suess, Dieter and Chumak, Andrii V.},	
    journal = {Nat. Electron.},
    title = {A universal inverse-design magnonic device},
    volume = {8},
    pages = {106-115},
    year = {2025},
	doi = {https://doi.org/10.1038/s41928-024-01333-7},
}

@article{Finocchio2024,
    author = {Finocchio, Giovanni and others},
    journal = {Nano Futures},
    title = {Roadmap for unconventional computing with nanotechnology},
    volume = {8},
    eid = {012001},
    year = {2024},
    doi = {10.1088/2399-1984/ad299a},
}

@article{Wang2024,
    author = {Wang, Qi and Csaba, Gyorgy and Verba, Roman and Chumak, Andrii V. and Pirro, Philipp},
    title = {Nanoscale magnonic networks},
    journal = {Phys. Rev. Appl.},
    volume = {21},
    eid = {040503},
    year = {2024},
    doi = {10.1103/PhysRevApplied.21.040503},
}

@article{Chumak2022,
    author={Chumak, A. V. and others},
    journal = {IEEE Trans. Magn.},
    title = {Advances in Magnetics Roadmap on Spin-Wave Computing},
    volume = {58},
    eid = {0800172},
    year = {2022},
    doi = {10.1109/TMAG.2022.3149664},
}

@article{Mahmoud2020,
    author = {Abdulqader Mahmoud and Florin Ciubotaru and Frederic Vanderveken and Andrii V. Chumak and Said Hamdioui and Christoph Adelmann and Sorin Cotofana},
    journal = {J. Appl. Phys.},
    title = {Introduction to spin wave computing},
    volume = {128},
    eid = {161101},
    year = {2020},
    doi = {10.1063/5.0019328},
}

@article{Kuznetsov2025,
    author = {Nikolai Kuznetsov and Huajun Qin and Lukáš Flajšman and Sebastiaan van Dijken},
    title = {Optical control of spin waves in hybrid magnonic- plasmonic structures},
    journal = {Sci. Adv.},
    volume = {11},
    eid = {eads2420},
    year = {2025},
    doi = {10.1126/sciadv.ads2420},
}

@article{Krcma2025,
    author = {Jakub Krčma and Ondrej Wojewoda and Martin Hrtoň and Jakub Holobrádek and Jon Ander Arregi and Jaganandha Panda and Ekaterina Pribytova and Michal Urbánek},
    title = {Mie-enhanced microfocused Brillouin light scattering for full wave vector resolution of nanoscale spin waves},
    journal = {Sci. Adv.},
    volume = {11},
    eid = {eady8833},
    year = {2025},
    doi = {10.1126/sciadv.ady8833},
}

@article{Heinz2020,
    author = {Björn Heinz and Thomas Brächer and Michael Schneider and Qi Wang and Bert Lägel and Anna M. Friedel and David Breitbach and Steffen Steinert and Thomas Meyer and Martin Kewenig and Carsten Dubs and Philipp Pirro and Andrii V. Chumak},
    title = {Propagation of spin-Wave packets in individual nanosized yttrium iron garnet magnonic conduits},
    journal = {Nano Lett.},
    volume = {20},
    pages = {4220-4227},
    year = {2020},
    doi = {10.1021/acs.nanolett.0c00657},
}

@article{Vogel2018,
    author = {Marc Vogel and Rick Aßmann and Philipp Pirro and Andrii V. Chumak and Burkard Hillebrands and Georg von Freymann},
    title = {Control of Spin-Wave Propagation using Magnetisation Gradients},
    journal = {Sci. Rep.},
    volume = {8},
    eid = {11099},
    year = {2018},
    doi = {10.1038/s41598-018-29191-2},
}

@article{Kargar2021,
    author = {Fariborz Kargar and Alexander A. Balandin},
    title = {Advances in Brillouin–Mandelstam light-scattering spectroscopy},
    journal = {Nat. Photon.},
    volume = {15},
    pages = {720-731},
    year = {2021},
    doi = {10.1038/s41566-021-00836-5},
}

@article{Sebastian2015,
    author = {Thomas Sebastian and Katrin Schultheiss and Björn Obry and Burkard Hillebrands and Helmut Schultheiss},
    title = {Micro-focused Brillouin light scattering: Imaging spin waves at the nanoscale},
    journal = {Front. Phys.},
    volume = {3},
    eid = {35},
    year = {2015},
    doi = {10.3389/fphy.2015.00035},
}

@article{Sandercock1973,
    author = {John Sandercock and Wolfram Wettling},
    title = {Light scattering from thermal acoustic magnons in yttrium iron garnet},
    journal = {Solid State Commun.},
    volume = {13},
    pages = {1729-1732},
    year = {1973},
    doi = {10.1016/0038-1098(73)90276-7},
}

@article{Bouvet2025,
    author = {Pierre Bouvet and Carlo Bevilacqua and Yogeshwari Ambekar and others},
    title = {Consensus statement on Brillouin light scattering microscopy of biological materials},
    journal = {Nat. Photon.},
    volume = {19},
    pages = {681-691},
    year = {2025},
    doi = {10.1038/s41566-025-01681-6},
}

@article{Antonacci2020,
    author = {Antonacci, G. and Beck, T. and Bilenca, A. and Czarske, J. and Elsayad, K. and Guck, J. and Kim, K. and Krug, B. and Palombo, F. and Prevedel, R. Scarcelli, G.},
    title = {Recent progress and current opinions in Brillouin microscopy for life science applications},
    journal = {Biophys. Rev.},
    volume = {12},
    pages = {615-624},
    year = {2020},
    doi = {10.1007/s12551-020-00701-9},
}

@article{Jiang2023,
    author = {Zhihao Jiang and Jinho Lim and Yi Li and Wolfgang Pfaff and Tzu Hsiang Lo and Jiangchao Qian and André Schleife and Jian Min Zuo and Valentine Novosad and Axel Hoffmann},
    journal = {Appl. Phys. Lett.},
    title = {Integrating magnons for quantum information},
    volume = {123},
    eid = {130501},
    year = {2023},
    doi = {10.1063/5.0157520},
}

@article{Li2020,
    author = {Yi Li and Wei Zhang and Vasyl Tyberkevych and Wai Kwong Kwok and Axel Hoffmann and Valentine Novosad},
    journal = {J. Appl. Phys.},
    title = {Hybrid magnonics: Physics, circuits, and applications for coherent information processing},
    volume = {128},
    eid = {130902},
    year = {2020},
    doi = {10.1063/5.0020277},
}

@article{Forsch2020,
    author = {Moritz Forsch and Robert Stockill and Andreas Wallucks and Igor Marinković and Claus Gärtner and Richard A. Norte and Frank van Otten and Andrea Fiore and Kartik Srinivasan and Simon Gröblacher},
    journal = {Nat. Phys.},
    title = {Microwave-to-optics conversion using a mechanical oscillator in its quantum ground state},
    volume = {16},
    pages = {69-74},
    year = {2020},
    doi = {10.1038/s41567-019-0673-7},
}

@article{Quirion2019,
    author = {Dany Lachance-Quirion and Yutaka Tabuchi and Arnaud Gloppe and Koji Usami and Yasunobu Nakamura},
    journal = {Appl. Phys. Express},
    title = {Hybrid quantum systems based on magnonics},
    volume = {12},
    eid = {070101},
    year = {2019},
    doi = {10.7567/1882-0786/ab248d},
}

@article{Krantz2019,
    author = {P. Krantz and M. Kjaergaard and F. Yan and T. P. Orlando and S. Gustavsson and W. D. Oliver},
    journal = {Appl. Phys. Rev.},
    title = {A quantum engineer's guide to superconducting qubits},
    volume = {6},
    eid = {021318},
    year = {2019},
    doi = {10.1063/1.5089550}
}

@article{Serha2025_YSGAG,
  author  = {Serha, R. O. and Dubs, C. and Guguschev, C. and Aichner, B. and Schmoll, David and Sch{\"a}fer, J. and Panda, J. and Weiler, M. and Pirro, P. and Urb{\'a}nek, M. and Chumak, A. V.},
  title   = {The ideal substrate for yttrium iron garnet films in quantum magnonics},
  journal = {Communications Materials},
  year    = {2026},
  volume  = {7},
  pages   = {134},
  doi     = {10.1038/s43246-026-01146-5}
}

@article{Serha2025_DampingEnhancement,
    author = {Serha, R. O. and Voronov, A. A. and Schmoll, D. and Klingbeil, R. and Knauer, S. and Koraltan, S. and Pribytova, E. and Lindner, M. and Reimann, T. and Dubs, C. and Abert, C. and Verba, R. and Urbánek, M. and Suess, D. and Chumak, A. V.},
    journal = {Materials Today Quantum},
    title = {Damping enhancement in {YIG} at millikelvin temperatures due to {GGG} substrate},
    volume = {5},
    eid = {100025},
    year = {2025},
    doi = {10.1016/j.mtquan.2025.100025},
}

@article{Schmoll2025_Wavenumber,
    author = {Schmoll, D. and Voronov, A. A. and Serha, R. O. and Slobodianiuk, D. and Levchenko, K. O. and Abert, C. and Knauer, S. and Suess, D. and Verba, R. and Chumak, A. V.},
    journal = {Phys. Rev. B},
    title = {Wavenumber-dependent magnetic losses in yttrium iron garnet–gadolinium gallium garnet heterostructures at millikelvin temperatures},
    volume = {111},
    eid = {134428},
    year = {2025},
    doi = {10.1103/PhysRevB.111.134428},
}

@article{Knauer_2023,
    author = {S. Knauer and K. Davídková and D. Schmoll and R. O. Serha and A. Voronov and Q. Wang and R. Verba and O. V. Dobrovolskiy and M. Lindnera and T. Reimann and C. Dubs and M. Urbánek and A. V. Chumak},
    journal = {J. Appl. Phys.},
    title = {Propagating spin-wave spectroscopy in a liquid-phase epitaxial nanometer-thick YIG film at millikelvin temperatures},
    volume = {133},
    eid = {143905},
    year = {2023},
    doi = {10.1063/5.0137437},
}

@article{Nakamura2020,
    author = {Dany Lachance-Quirion and Samuel Piotr Wolski and Yutaka Tabuchi and Shingo Kono and Koji Usami and Yasunobu Nakamura},
    journal = {Science},
    title = {Entanglement-based single-shot detection of a single magnon with a superconducting qubit},
    volume = {367},
    pages = {425-428},
    year = {2020},
    doi = {10.1126/science.aaz9236},
}

@article{Tabuchi2015,
    author = {Yutaka Tabuchi and Seiichiro Ishino and Atsushi Noguchi and Toyofumi Ishikawa and Rekishu Yamazaki and Koji Usami and Yasunobu Nakamura},
    journal = {Science},
    title = {Coherent coupling between a ferromagnetic magnon and a superconducting qubit},
    volume = {349},
    pages = {405-408},
    year = {2015},
    doi = {10.1126/science.aaa3693}
}

@incollection{BHOI_book,
author = {Biswanath Bhoi and Sang-Koog Kim},
booktitle = {Solid State Physics},
title = {Chapter Two - Roadmap for photon-magnon coupling and its applications},
editor = {Robert L. Stamps},
publisher = {Academic Press},
volume = {71},
pages = {39-71},
year = {2020},
issn = {0081-1947},
doi = {10.1016/bs.ssp.2020.09.004}
}

@incollection{BHOI_book2,
author = {Biswanath Bhoi and Sang-Koog Kim},
booktitle = {Solid State Physics},
title = {Chapter One - Photon-magnon coupling: Historical perspective, status, and future directions},
editor = {Robert L. Stamps and Helmut Schultheiß},
publisher = {Academic Press},
volume = {70},
pages = {1-77},
year = {2019},
issn = {0081-1947},
doi = {10.1016/bs.ssp.2019.09.001}
}

@misc{supplement,
  author = {Schmoll, D. and Kuznetsov, N. and Rehberger, P. and Vilsmeier, F. and Verba, R. and Slobodianiuk, D. and Serha, R. O. and Levchenko, K. O. and van Dijken, S. and Chumak, A. V. and Knauer, S.},
  title = {{Supplemental} {Material} to \textit{{Brillouin} Light Scattering Spectroscopy of Propagating Magnons at Sub-Kelvin Temperatures},},
  year   = {2026}
}

@article{Falkner2025_BrillouinCancer,
    author = {Falkner, N. T. and Duman, Meryem-Nur and Zabolizadeh, Z. and Mahmodi, H. and Shi, C. and Zhang, J. Cox, T. R. and Kabakova, I. V.},
    title = {Brillouin microscopy in cancer research: a review},
    journal = {Journal of Biomedical Optics},
    volume = {30},
    number = {12},
    pages = {124509},
    year = {2025},
    doi = {10.1117/1.JBO.30.12.124509}
}

@article{Bevilacqua2023_LineScanBrillouin,
    author = {Bevilacqua, C. and Gomez, J. M. and Fiuza, Ulla-Maj and Chan, C. J. and Wang, L. and Hambura, S. and Eguren, M. and Ellenberg, J. and Diz-Muñoz, A. and Leptin, M. and Prevedel, R.},
    title = {High-resolution line-scan {Brillouin} microscopy for live imaging of mechanical properties during embryo development},
    journal = {Nature Methods},
    volume = {20},
    pages = {755--764},
    year = {2023},
    doi = {10.1038/s41592-023-01822-1}
}

@article{Keshmiri2024_BrillouinAnisotropy,
    author = {Keshmiri, H. and Cikes, D. and Samalova, M. and Schindler, L. and Appel, Lisa-Marie and Urbanek, M. and Yudushkin, I. and Slade, D. and Weninger, W. J. and Peaucelle, A. and Penninger, J. and Elsayad, K.},
    title = {Brillouin light scattering anisotropy microscopy for imaging the viscoelastic anisotropy in living cells},
    journal = {Nature Photonics},
    volume = {18},
    pages = {276--285},
    year = {2024},
    doi = {10.1038/s41566-023-01368-w}
}

@article{Kimble2008_QuantumInternet,
    author = {Kimble, H. J.},
    title = {The quantum internet},
    journal = {Nature},
    volume = {453},
    pages = {1023--1030},
    year = {2008},
    doi = {10.1038/nature07127}
}

@article{Reiserer2015,
  title = {Cavity-based quantum networks with single atoms and optical photons},
  author = {Reiserer, Andreas and Rempe, Gerhard},
  journal = {Rev. Mod. Phys.},
  volume = {87},
  pages = {1379--1418},
  year = {2015},
  doi = {10.1103/RevModPhys.87.1379}
}

@article{Wehner2018_QuantumInternet,
    author = {Wehner, S. and Elkouss, D. and Hanson, R.},
    title = {Quantum internet: A vision for the road ahead},
    journal = {Science},
    volume = {362},
    pages = {eaam9288},
    year = {2018},
    doi = {10.1126/science.aam9288}
}

@article{Boona2014_PRB,
    author = {Boona, Stephen R and Heremans, Joseph P},
    title = {Magnon thermal mean free path in yttrium iron garnet},
    journal = {Physical Review B},
    volume = {90},
    pages = {064421},
    year = {2014},
    doi = {10.1103/PhysRevB.90.064421}
}

@article{bhandari1966scattering,
    author = {Bhandari, CM and Verma, GS},
    title = {Scattering of magnons and phonons in the thermal conductivity of yttrium iron garnet},
    journal = {Physical Review},
    volume = {152},
    pages = {731},
    year = {1966},
    doi = {10.1103/PhysRev.152.731}
}

@article{walton1973thermal,
  author={Walton, D and Rives, JE and Khalid, Q},
  title={Thermal transport by coupled magnons and phonons in yttrium iron garnet at low temperatures},
  journal={Physical Review B},
  volume={8},
  pages={1210},
  year={1973},
  doi = {10.1103/PhysRevB.8.1210}
}

@article{pan2013specific,
  author={Pan, BY and Guan, TY and Hong, XC and Zhou, SY and Qiu, X and Zhang, H and Li, SY},
  title={Specific heat and thermal conductivity of ferromagnetic magnons in Yttrium Iron Garnet},
  journal={Europhysics Letters},
  volume={103},
  pages={37005},
  year={2013},
  doi = {10.1209/0295-5075/103/37005}
}

@article{daudin1982thermodynamic,
  author={Daudin, B and Lagnier, R and Salce, B},
  title={Thermodynamic properties of the gadolinium gallium garnet, {$\mathrm{Gd}_3\mathrm{Ga}_5\mathrm{O}_{12}$}, between 0.05 and 25 {K}},
  journal={Journal of Magnetism and Magnetic Materials},
  volume={27},
  pages={315--322},
  year={1982},
  doi = {10.1016/0304-8853(82)90092-0}
}

@article{prakash2018evidence,
  author={Prakash, Arati and Flebus, Benedetta and Brangham, Jack and Yang, Fengyuan and Tserkovnyak, Yaroslav and Heremans, Joseph P},
  title={Evidence for the role of the magnon energy relaxation length in the spin Seebeck effect},
  journal={Physical Review B},
  volume={97},
  pages={020408},
  year={2018},
  doi = {10.1103/PhysRevB.97.020408}
}

@article{simon1994_AlN,
  author={Simon, NJ},
  title={Cryogenic properties of inorganic insulation materials for ITER magnets: A review},
  journal={NIST},
  volume={5030},
  year={1994},
  publisher={National Institute of Standards and Technology, Boulder, CO (US)}
}

@article{Serha2026,
  author  = {Serha, R. O. and Mcallister, K. H. and Majcen, F. and Knauer, S. and Reimann, T. and Dubs, C. and Melkov, G. A. and Serga, A. A. and Tyberkevych, V. S. and Chumak, A. V. and Bozhko, D. A.},
  title   = {Ultralong-living magnons in the quantum limit},
  journal = {Science Advances},
  volume  = {12},
  number  = {18},
  pages   = {eaee2344},
  year    = {2026},
  doi     = {10.1126/sciadv.aee2344}
}

@article{Serha2026_APL,
  author  = {Serha, R. O. and Dubs, C. and Chumak, A. V.},
  title   = {Magnetic materials for quantum magnonics},
  journal = {APL Materials},
  volume  = {14},
  number  = {3},
  pages   = {030901},
  year    = {2026},
  doi     = {10.1063/5.0306423}
}

@article{vanLoo2018,
  author  = {van Loo, A. F. and Morris, R. G. E. and Karenowska, A. D.},
  title   = {Time-resolved measurements of surface spin-wave pulses at millikelvin temperatures},
  journal = {Phys. Rev. Applied},
  volume  = {10},
  pages   = {044070},
  year    = {2018},
  doi     = {10.1103/PhysRevApplied.10.044070}
}

@ARTICLE{Demokritov2008,
  author={Demokritov, Sergej O. and Demidov, Vladislav E.},
  journal={IEEE Transactions on Magnetics}, 
  title={Micro-Brillouin Light Scattering Spectroscopy of Magnetic Nanostructures}, 
  year={2008},
  volume={44},
  pages={6-12},
  doi={10.1109/TMAG.2007.910227}
  }

@article{Demokritov1986,
  author  = {Demokritov, S. O. and Kre{\u{\i}}nes, N. M. and Kudinov, V. I.},
  title   = {Scattering of light by magnons in two branches of the spectrum of antiferromagnetic {EuTe}},
  journal = {Pis'ma Zh. Eksp. Teor. Fiz.},
  volume  = {43},
  number  = {6},
  pages   = {312--314},
  year    = {1986}
}

@article{Demokritov1987,
  author  = {Demokritov, S. O. and Kre{\u{\i}}nes, N. M. and Kudinov, V. I.},
  title   = {Inelastic scattering of light in the antiferromagnet {EuTe}},
  journal = {Zh. Eksp. Teor. Fiz.},
  volume  = {92},
  pages   = {689--703},
  year    = {1987}
}

@article{Che2025,
  author  = {Che, P. and Ciola, R. and Garst, M. and Kravchuk, V. and Baral, P. R. and Magrez, A. and Berger, H. and Schönenberger, T. and Rønnow, H. M. and Grundler, D.},
  title   = {Short-wave magnons with multipole spin precession detected in the topological bands of a skyrmion lattice},
  journal = {Communications Materials},
  volume  = {6},
  pages   = {139},
  year    = {2025},
  doi     = {10.1038/s43246-025-00858-4}
}

@article{Vogel2015,
  author  = {Vogel, M. and Chumak, A. V. and Waller, E. H. and Langner, T. and Vasyuchka, V. I. and Hillebrands, B. and von Freymann, G.},
  title   = {Optically reconfigurable magnetic materials},
  journal = {Nature Physics},
  volume  = {11},
  pages   = {487--491},
  year    = {2015},
  doi     = {10.1038/nphys3325}
}


\clearpage
\onecolumngrid

\setcounter{section}{0}
\setcounter{subsection}{0}
\setcounter{figure}{0}
\setcounter{table}{0}
\setcounter{equation}{0}

\renewcommand{\thesection}{\Roman{section}}      
\renewcommand{\thesubsection}{\Alph{subsection}} 
\renewcommand{\thefigure}{S\arabic{figure}}      
\renewcommand{\thetable}{S\arabic{table}}        
\renewcommand{\theequation}{S\arabic{equation}}  

\begin{center}
    \vspace{1cm}
    \Large \textbf{Supplementary Information:  Brillouin Light Scattering Spectroscopy of
Propagating Magnons at Sub-Kelvin Temperatures}
    \vspace{0.5cm}
\end{center}

\section{Detailed Description of the Dilution Refrigerator Coupled Brillouin Light Scattering Spectroscopy Setup}

\begin{figure}[t]
    \centering
    \includegraphics[width=1\textwidth]{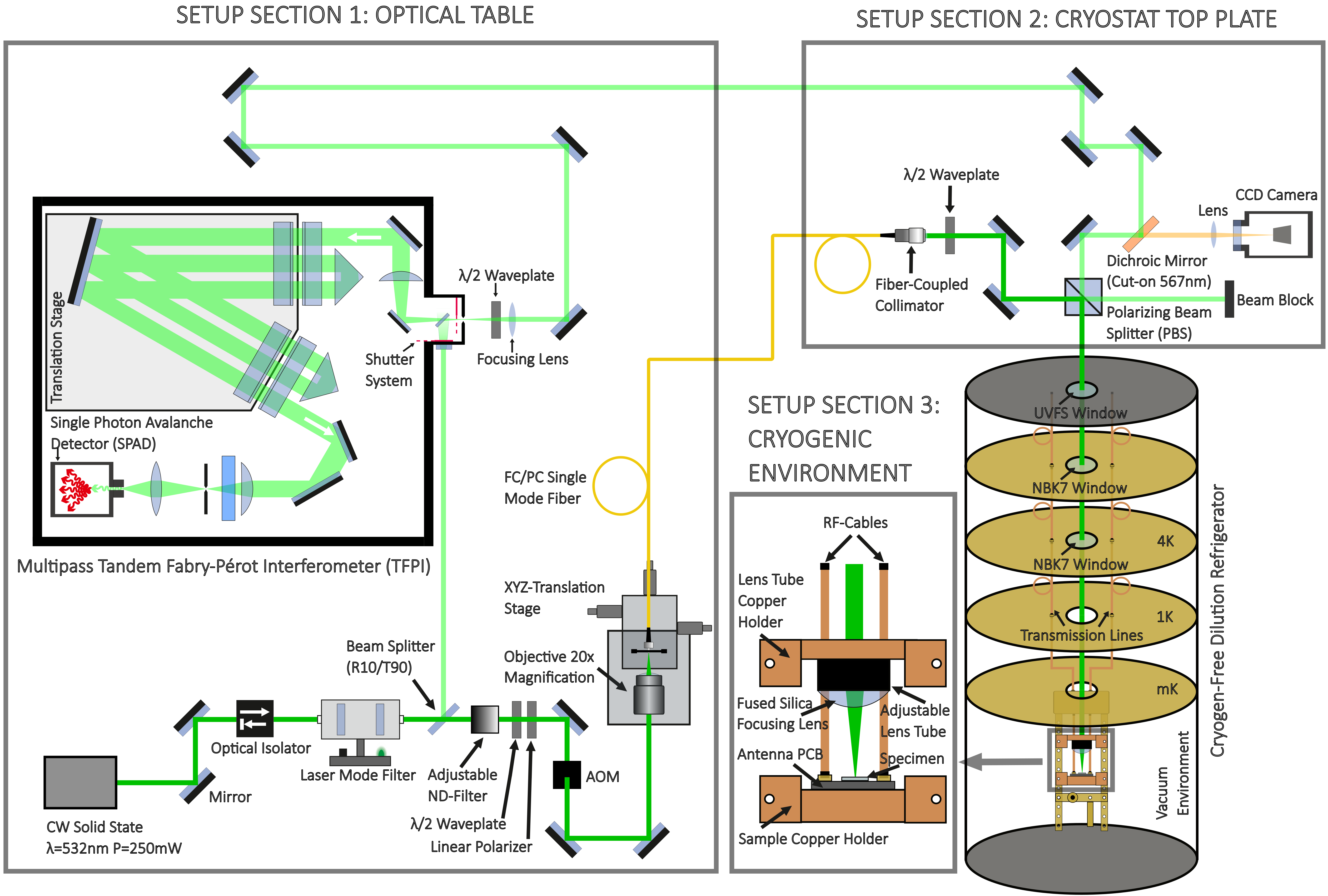}
    \captionsetup{justification=justified}
    \caption{\justifying Schematic illustration of the beam path and optical components of the dilution refrigerator coupled BLS assembly. The experimental setup is structured into three sections: the optical table, the cryostat top plate, and the cryogenic environment.}
    \label{fig:CryoBLS}
\end{figure}

A classic Brillouin Light Scattering (BLS) spectroscopy setup is coupled with a Bluefors LD250 cryogen-free dilution refrigerator system for the optical detection of magnons at cryogenic temperatures. The experimental setup is schematically illustrated in Fig.~\ref{fig:CryoBLS} and can be split into three main sections: an optical table at room temperature, the cryostat top plate, and the cryogenic environment.

A continuous wave, diode pumped solid-state (DPSS) laser (Novanta Torus 532 250 mpc, $\lambda=$ \qty{532}{\nano\meter}, $L_\mathrm{P}=$~\qty{250}{\milli\W}) serves as the coherent light source. The laser is protected against back-reflection from subsequent components via an optical isolator. Spurious secondary laser modes are suppressed via a temperature-stabilised etalon filter (Table Stable TCF-2), which is followed by a plate beam splitter that separates $10\%$ of the light intensity as a reference beam towards the Tandem Fabry Pérot Interferometer (TFPI). The remaining $90\%$ of the laser power can be regulated using a neutral-density filter wheel with variable optical density. 

To minimise the average thermal load introduced into the dilution refrigerator, the beam can be pulsed via an Acousto-Optic Modulator (AOM)~[1]. The AOM (model 3080-125 from G\&H) utilises a $TeO_2$ crystal, driven by a transducer (model 1080AFP-AD-1.0-EC80 from G\&H) with an RF-signal centred at a frequency of \qty{80}{\mega\hertz} to modulate the length and delay of the pulses with a minimum rise time of \qty{25}{\nano\second}. To optimise the diffraction efficiency, a $\lambda /2$-waveplate and a linear polariser are used to define the polarisation state of the light before it enters the crystal. The first-order diffracted beam is subsequently focused by an objective lens mounted on an xyz-translation stage into the core of a \qty{10}{\meter}-long single-mode (SM) optical fibre (type 460HP). This fibre is terminated with an FC/APC-connector at the end of the optical table to minimise optical feedback. The SM-fibre guides light to the second section of the setup, located on top of the dilution refrigerator.

In addition to the components required for preparing the incident beam, the optical table houses the TFPI and a Single Photon Avalanche Detector (SPAD). The backscattered light is guided from the top plate of the dilution refrigerator towards the interferometer as a free-space beam. An achromatic-doublet lens with a focal length of \qty{100}{\milli\meter} focuses the light into the input pinhole of the TFPI, and a $\lambda/2$-waveplate allows for rotating the polarisation state of the light, as the internal optics of the interferometer model are optimised for vertical polarization.

The components in the second setup section are mounted on an optical breadboard that is fixed to the top plate of the cryostat. To mitigate vibrations, the dilution refrigerator, and consequently the breadboard, are decoupled from the static frame by an active vibration isolation system (Table Stable AVI-200 LP). From the optical fibre, the incident light is re-collimated into a free-space beam and directed toward a polarising beam-splitter (PBS) cube, which reflects the majority of the light into the dilution refrigerator. A $\lambda/2$-waveplate, positioned before the PBS, provides additional control over the laser intensity directed toward the cryostat. The fraction of light that is inelastically scattered by spin waves undergoes a $90^\circ$ polarization rotation and consequently is transmitted by the PBS after it exits the cryostat again. The elastically scattered light and the majority of the light inelastically scattered from phonons, on the other hand, retain the incident polarisation and are reflected toward a beam dump. The backscattered light, which is transmitted through the PBS, is subsequently guided by a mirror assembly back to the optical table for analysis.

Optical access into the dilution refrigerator is provided via a \qty{4}{\milli\meter}-thick Ultra Violet Fused Silica (UVFS) window, Anti-Reflection (AR) coated for the 350-\qty{700}{\nano\meter} wavelength range. To suppress thermal black-body radiation and maintain the system's base temperature, two additional \qty{1}{\milli\meter}-thick AR-coated N-BK7 windows are mounted at the \qty{50}{\kelvin} and \qty{4}{\kelvin} stages inside the vacuum environment. An AR-coated, 1-inch plano-convex UVFS-lens with a focal length of $f=$~\qty{35}{\milli\meter} focuses the incident beam onto the investigated sample and re-collimates the backscattered light back toward the PBS. The lens is housed within an adjustable, vacuum-compatible lens tube, which is threaded into an Oxygen-Free High-Conductivity (OFHC) copper adapter and secured with a retaining ring. The adapter is mechanically connected to the cryostat's sample holder puck and thermally anchored to the mixing chamber stage. Given that the sample position is fixed within the cryostat, focal adjustment must be performed ex situ by modulating the lens tube position before cool-down. 

\section{Description of the Numerical Model for Local Laser Heating}

Laser-induced local heating during BLS spectroscopy performed in a dilution refrigerator was modelled by integrating three-dimensional (3D) optical and thermal simulations.

\subsection{Optical simulations}

The spatial distribution of the absorbed optical power was calculated using the finite-difference time-domain (FDTD) solver in Ansys Lumerical. The simulated geometry consisted of an \qty{18}{\micro\m}-thick YIG film on a GGG substrate. The three-dimensional computational domain measured $15 \times 15 \times 45$ ~\qty{}{\micro\m^3}. Since GGG is optically transparent at the probe-laser wavelength of \qty{532}{\nm}, the substrate thickness was reduced from its physical value of \qty{500}{\micro\m} to \qty{25}{\micro\m} to reduce computational cost while preserving the optical field distribution within both layers. The complex refractive indices of YIG and GGG were determined experimentally: the real parts were obtained by ellipsometry, and the extinction coefficients from optical transmittance and reflectance measurements. The \qty{500}{\micro\m}-thick AlN substrate and the \qty{18}{\micro\m}-thick Cu metallisation layer were excluded from the optical model, as optical absorption is confined predominantly to the YIG layer and incident light does not penetrate appreciably into the underlying structure. A Gaussian beam with a wavelength of \qty{532}{\nano\m} and a waist diameter of approximately \qty{5}{\micro\m} was focused through the GGG substrate onto the YIG layer, reproducing the experimental back-scattering BLS geometry. Incident optical powers were set to match the experimental conditions. Perfectly matched layer (PML) boundary conditions were applied on all sides of the computational domain to suppress spurious reflections. The volumetric absorbed power density within the YIG layer was extracted using a built-in advanced power absorption monitor and subsequently used as the heat source distribution in the thermal simulations.

\subsection{Thermal simulations}

\begin{figure}[t]
    \centering
    \includegraphics[width=1\textwidth]{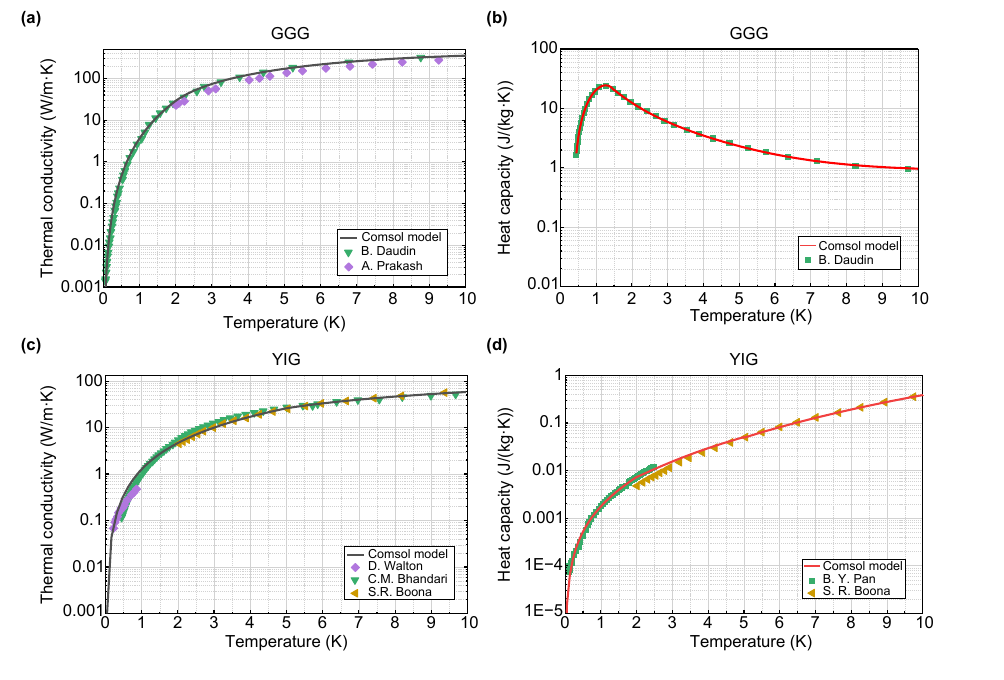}
    \captionsetup{justification=justified}
    \caption{\justifying Temperature-dependent thermal conductivity and heat capacity of \textbf{(a,b)} GGG and \textbf{(c,d)} YIG, respectively. Symbols represent experimental data from literature; solid lines show the corresponding fits used in the thermal simulations. The thermal conductivity of YIG includes both phonon and magnon contributions and was taken from literature data measured at zero applied magnetic field.}
    \label{fig:Thermal Properties}
\end{figure}

Thermal simulations were performed in COMSOL Multiphysics using both steady-state and transient heat transfer in solids modules. The computational domain had lateral dimensions of $400 \times 400 $ ~\qty{}{\micro\m^2} and comprised the \qty{500}{\micro\m}-thick GGG substrate, the \qty{18}{\micro\m}-thick YIG film, the \qty{500}{\micro\m}-thick AlN substrate, and the \qty{18}{\micro\m}-thick Cu layer. Temperature-dependent thermal conductivities and heat capacities were implemented by fitting experimentally reported low-temperature data from the literature wherever available (Refs.~[2-8]). For materials whose reported data did not span the full temperature range considered here, the relevant properties were extrapolated from the lowest measured values available. Temperature-dependent thermal properties for YIG and GGG were implemented over the complete simulation temperature range; these are summarised in Fig.~\ref{fig:Thermal Properties}.

Steady-state simulations were performed to determine the equilibrium temperature distribution under continuous-wave optical illumination. A fixed-temperature boundary condition was applied to the bottom surface of the Cu layer, set equal to the experimentally measured global temperature after thermal equilibration of the system. All remaining external boundaries were treated as thermally insulating.

In the transient simulations, the thermal coupling between the modelled structure and the dilution refrigerator was modelled using a thermal resistance boundary condition applied to the bottom surface of the Cu layer. The boundary heat flux was defined as $q = h(T_\mathrm{p}^\mathrm{global} - T),$ where $h$ is an effective heat transfer coefficient and $T_\mathrm{p}^\mathrm{global}$ is the experimentally measured equilibrium temperature of the dilution refrigerator's sample puck via a ruthenium oxide resistance thermometer. This formulation accounts for the finite thermal conductance between the sample stack and the cryostat, without requiring explicit modelling of the complete thermal path — including mechanical interfaces, contact pressures, and interfacial thermal resistances — all of which are poorly constrained. These contributions are therefore subsumed into the single empirical parameter $h$. Its value, $h = 6170$ W/(m$^2$$\cdot$K), was estimated from the absorbed optical power and the experimentally measured temperature rise under pulsed laser illumination. The pulsed heat source was implemented via the Events interface, with a pulse duration of \qty{1}{\micro\s} and a period of \qty{4}{\micro\s}. In the transient regime, the extremely small heat capacity of YIG at millikelvin temperatures renders the system of differential equations numerically stiff, substantially increasing computational cost. To ensure numerical stability, the specific heat capacity was bounded below at 0.001 J/(kg$\cdot$K).

\section{Numerically Modelled Local Temperature Distribution for the Comparison of Optical and Electrical Transmission Spectra}
\begin{figure}[h]
    \centering
    \includegraphics[width=1\textwidth]{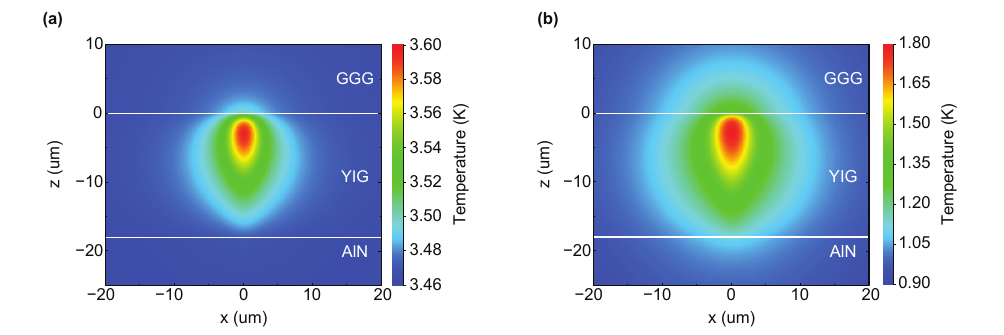}
    \captionsetup{justification=justified}
    \caption{\justifying Simulated steady-state temperature distribution in the vicinity of the laser focal spot for the GGG/YIG/AlN stack. The incident optical power at the YIG layer was $(79\pm2)$~\qty{}{\micro\W}, with the bottom boundary temperature fixed at \textbf{(a)} 3.46 K and \textbf{(b)} 540 mK.}
    \label{fig:Thermal modeling}
\end{figure}

\section{Magnon Population Analysis and Route to the Single-Magnon Regime}

To benchmark how close the presented BLS platform is to the goal of resolving individual magnons, and to identify the most efficient route towards it, we estimate both the thermal magnon population and the population of coherently excited magnons under our experimental conditions.

\subsection{Theoretical Framework and Parameters}

The occupation of a single magnon level of energy $E_\mathrm{k} = \hbar\omega_\mathrm{k}$ follows the Bose–Einstein distribution,

\begin{equation}
	\bar n(\omega_\mathrm{k}, T) = \left(\exp\left[\frac{\hbar \omega_\mathrm{k}}{k_\mathrm{B} T}\right] -1 \right)^{-1} .
\end{equation}

The total thermal magnon density is obtained by integrating the occupation rate weighted by the density of states over the magnon spectrum. For the comparatively thick film studied here, the three-dimensional density of states applies and, within the exchange approximation of the spin-wave spectrum $\omega_\mathrm{k} =  \omega_0 + D k^2$ with $D = \omega_\mathrm{M} \lambda_\mathrm{ex}^2$, the thermal magnon density reads
\begin{equation}\label{e:thermal-all}
	n_\mathrm{th}(T) = \frac1{4\pi^2}\int\limits_{\omega_0}^{\infty} \frac{\sqrt{\omega-\omega_0}}{D^{3/2}} \bar n(\omega,T) d\omega .
\end{equation} 

In principle, the integration should extend to the highest magnon frequency of the first Brillouin zone, and account for the deviation of the spectrum from the quadratic law near the zone boundary. At the low temperatures of interest here (below \qty{10}{\kelvin}), however, the thermal population is dominated by low-energy magnons, so that the closed form of Eq.~\ref{e:thermal-all} is highly accurate and fully sufficient. The same expression yields the thermal magnon number within any chosen frequency band by restricting the integration accordingly, enabling to evaluate the populations relevant to the BLS detection bandwidth and to the magnon linewidth only. The density of coherently excited magnons follows from the spin-wave power. Using the energy density $W = n\hbar\omega$, the spin-wave power $P_\mathrm{SW} = WSv_\mathrm{gr}$, where $S$ is the film cross-section and $v_\mathrm{gr}$ the group velocity, and accounting for the fraction $\eta$ of the input microwave power $P$ that is converted into spin waves ($P_\mathrm{SW} = \eta P$), the density of coherently excited magnons is
\begin{equation}
	n_\mathrm{coh} = \frac{\eta P}{ \hbar \omega v_\mathrm{gr} t w} ,
\end{equation} 

\noindent
with $t$ and $w$ as the film's thickness and width. The power transfer efficiency is estimated from the experimentally obtained $S_{21}$ transmission parameter. Assuming that the excitation and detection rate are the same, $S_{21} = \eta^2 \exp[-2\Gamma L/v_\mathrm{gr}]$, where $L$ is the distance between the antennas and $\Gamma \approx \alpha_\mathrm{eff} \omega$ is the magnon dissipation rate.

In the following calculations, we use $\mu_0 M_\mathrm{s}~=$~\qty{0.246}{\tesla}~[9] and $A_\mathrm{ex}~=$~\qty{5}{\pico\joule\meter^{-1}}~[10] for the YIG sample below \qty{10}{\kelvin}. The effective field was corrected for the contribution of the GGG stray field following the procedure of~[11], to match the FMR frequency of \qty{7.75}{\giga\hertz}. The effective damping parameter is assumed to be $\alpha_\mathrm{eff} = 10^{-3}$, as it was reported for micrometer-thick YIG films at temperatures below \qty{1}{\kelvin}~[12], and the group velocity is evaluated in the long-wavelength limit as $v_\mathrm{gr} = \omega_\mathrm{M}^2t/(4\omega_0)$. The film width is $w~=$~\qty{2}{\milli\meter}, the BLS detection linewidth is taken as \qty{400}{\mega\hertz}, and the BLS focal spot is approximately \qty{5}{\micro\meter} in diameter.

\subsection{Thermal and Coherent Magnon Population}

\begin{figure}[h]
    \centering
    \includegraphics[width=1\textwidth]{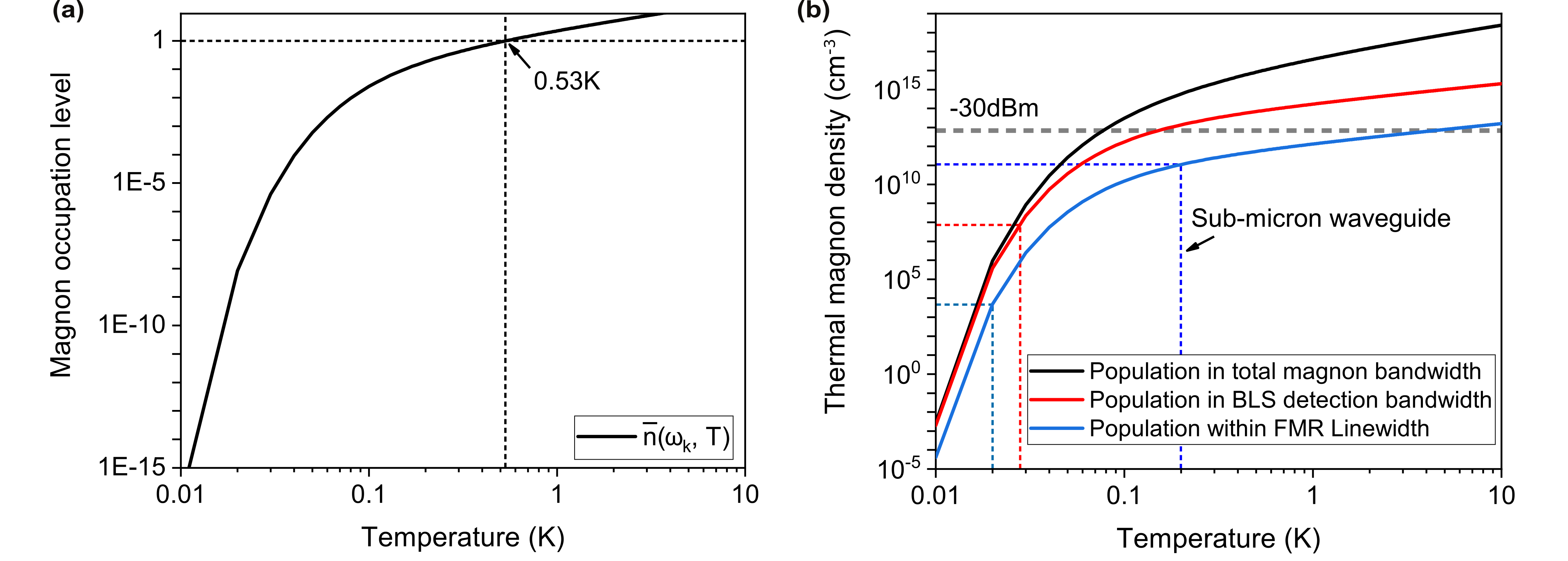}
    \captionsetup{justification=justified}
    \caption{\justifying \textbf{(a)} Occupation $\bar n(\omega_\mathrm{k}, T)$ of a single magnon level near the ferromagnetic-resonance frequency as a function of temperature (double-logarithmic scale). The dashed line marks the single-magnon level occupation $\bar n(\omega_\mathrm{k}, T)=1$ and the corresponding temperature. \textbf{(b)} Thermal magnon density as a function of temperature for the total magnon population (black), the population within the BLS detection bandwidth (red), and the population within the FMR linewidth (blue). The dashed red and blue lines indicate the single-magnon condition for the BLS and coherent-transmission experiments, respectively, with the dark blue dashed line indicating the corresponding condition for a nanoscale waveguide. The horizontal grey line marks the coherently excited magnon population at an applied microwave power of -\qty{30}{\dBm}.}
    \label{fig:n-level}
\end{figure}

Figure~\ref{fig:n-level}~(a) depicts the occupation of a single magnon level near the ferromagnetic-resonance (FMR) frequency as a function of temperature. The single-magnon level occupation, $\bar n(\omega_\mathrm{k}, T)<1$, is reached below \qty{0.53}{\kelvin}. Notably, this threshold is very close to the temperatures already attained in the presented experiments.

Because every magnon level has a finite linewidth set by the magnon lifetime, the most stringent benchmark for single-magnon operation is the magnon number integrated over
the relevant detection volume and bandwidth, rather than the occupation of an isolated level.  Figure~\ref{fig:n-level}~(b) illustrates the full thermal magnon density for the total magnon population (black curve), the populations contained within the BLS detection bandwidth (red curve), and the population within the magnon (FMR) linewidth (blue). For the \qty{18}{\micro\meter}-thick YIG film used in the experiments and a laser focal spot of \qty{5}{\micro\meter}, the volume-integrated single-magnon condition $N_\mathrm{th} = n_\mathrm{th}V < 1$ for the BLS detection bandwidth corresponds to a temperature of $\approx$~\qty{28}{\milli\kelvin} (red dashed line in Fig.~\ref{fig:n-level}~(b)). However, this value is not a fundamental limit, it reflects the large detection volume of the thick film and can therefore be shifted by reducing the film thickness. For example, at a waveguide width of \qty{100}{\nano\meter} the single magnon condition corresponds already to a temperature increased by a factor of two. 

The same picture applies to coherent-transmission experiments, for which the single-magnon
condition corresponds to a magnon level linewidth ($\approx 2\alpha_\mathrm{eff}\omega$) and yields a temperature of $\approx$~\qty{20}{\milli\kelvin} (blue dashed line in Fig.~\ref{fig:n-level}~(b)) for the large-area film of the experiments. For a typical magnonic conduit of \qty{100}{\nano\meter} thickness, \qty{1}{\micro\meter} width, and \qty{100}{\micro\meter} length, the condition $N < 1$ is already satisfied at \qty{200}{\milli\kelvin} (dark blue dashed line in Fig.~\ref{fig:n-level}~(b)).

Besides reducing the sample volume, another effective strategy is to operate at higher magnon frequencies. Since increasing the FMR frequency $\omega_0$ by a factor $p$ raises all characteristic temperatures by the same factor, working with \qty{50}{\giga\hertz} magnons, for example, would shift the single-magnon BLS condition to $\approx$~\qty{0.3}{\kelvin} and to $\approx$~\qty{0.6}{\kelvin} when combined with a reduced film thickness. Both targets lie within the operating range of standard dilution refrigerators, underscoring that the single-magnon regime is within practical reach. 

\begin{figure}[h]
    \centering
    \includegraphics[width=0.5\textwidth]{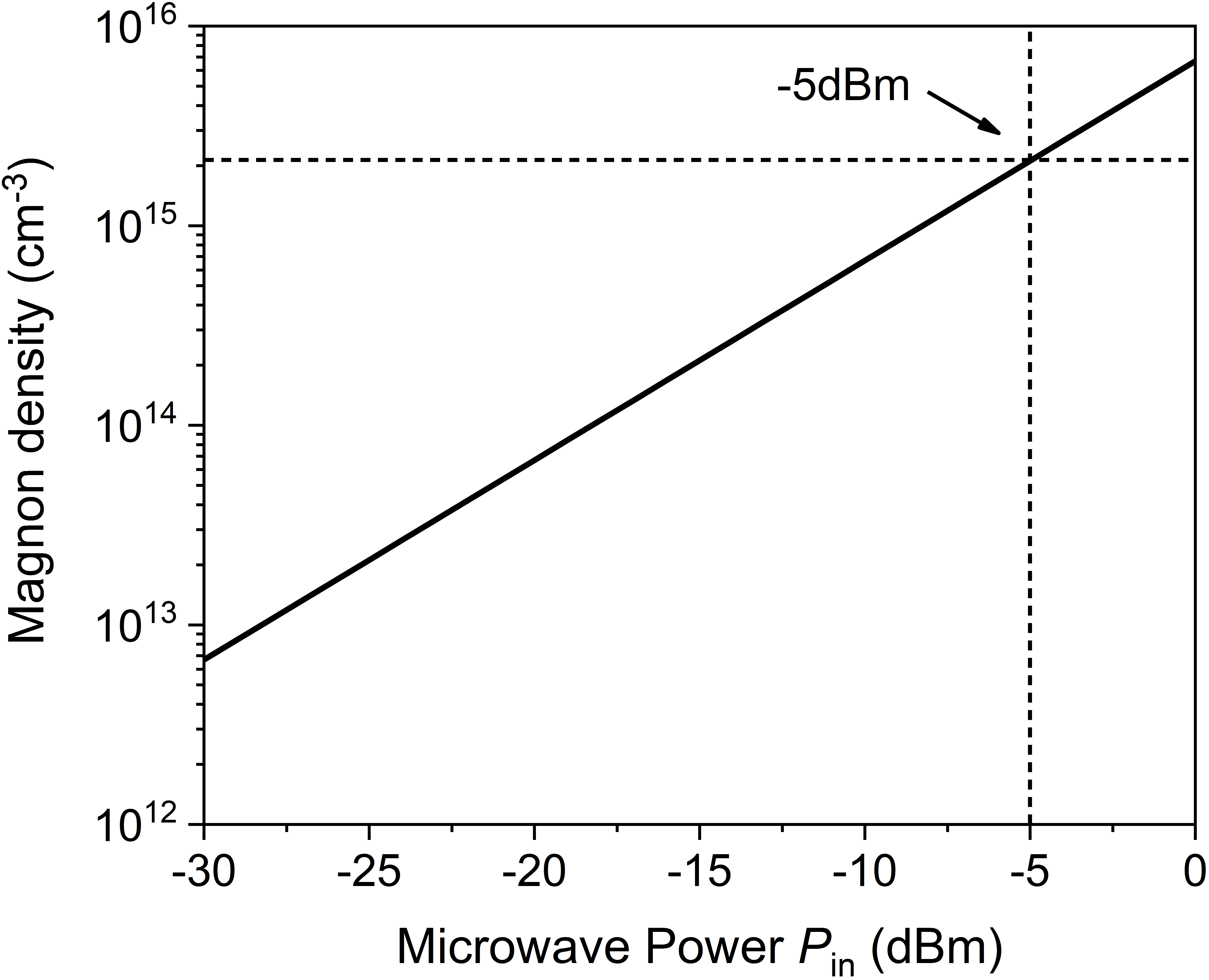}
    \captionsetup{justification=justified}
    \caption{\justifying Density of coherently excited magnons as a function of the applied microwave power. The black dashed lines highlight the microwave drive used for the BLS spectroscopy measurements with a continuous feed of optical power.}
    \label{fig:powerdependence}
\end{figure}

Finally, Fig.~\ref{fig:powerdependence} shows the density of coherently excited magnons as a function of the applied microwave power at the antenna $P_\mathrm{in}$. The power of \qty{-5}{\dBm}, used for the BLS spectroscopy measurements with a continuous feed of optical power, is highlighted via black dashed lines. Even at a drive of \qty{-30}{\dBm}, the coherently excited population exceeds the thermal magnon population within the coherent magnon linewidth for all temperatures below \qty{4}{\kelvin}. The coherent signal of interest therefore stands well above the thermal background throughout the temperature range of the presented experiments, ensuring a favourable signal contrast for coherent magnon detection.
Taken together, these estimates show that the present platform does not yet meet the single-magnon level limit due to the increased temperature induced by laser heating and the necessary microwave drive of coherent magnons. However, the remaining step towards optical single-magnon resolution is in reach of the experiments. A decrease of the local temperature $T_\mathrm{p}^\mathrm{local}$, combined with nanoscale magnonic conduits, and higher spin-wave frequencies, represents a well-defined and experimentally established route to Brillouin light scattering spectroscopy in the single-magnon regime.

\clearpage

\section{Supplementary References}
\begin{enumerate}[label={[\arabic*]}, leftmargin=3em, labelsep=0.5em]
    \item M. R. Schweizer, F. Kühn, V. S. L’vov, A. Pomyalov, G. von Freymann, B. Hillebrands, and A. A. Serga, \textit{Applied Physics Letters} \textbf{124}, 092402 (2024).
    \item S. R. Boona and J. P. Heremans, \textit{Physical Review B} \textbf{90}, 064421 (2014).
    \item C. Bhandari and G. Verma, \textit{Physical Review} \textbf{152}, 731 (1966).
    \item D. Walton, J. Rives, and Q. Khalid, \textit{Physical Review B} \textbf{8}, 1210 (1973).
    \item B. Pan, T. Guan, X. Hong, S. Zhou, X. Qiu, H. Zhang, and S. Li, \textit{Europhysics Letters} \textbf{103}, 37005 (2013).
    \item B. Daudin, R. Lagnier, and B. Salce, \textit{Journal of Magnetism and Magnetic Materials} \textbf{27}, 315 (1982).
    \item A. Prakash, B. Flebus, J. Brangham, F. Yang, Y. Tserkovnyak, and J. P. Heremans, \textit{Physical Review B} \textbf{97}, 020408 (2018).
    \item N. Simon, NIST 5030 (1994).
    \item D. Schmoll, A. A. Voronov, R. O. Serha, D. Slobodianiuk, K. O. Levchenko, C. Abert, S. Knauer, D. Suess, R. Verba, and A. V. Chumak, \textit{Phys. Rev. B} \textbf{111}, 134428 (2025).
    \item R. C. LeCraw and L. R. Walker, \textit{Journal of Applied Physics} \textbf{32}, S167 (1961).
    \item R. O. Serha, A. A. Voronov, D. Schmoll, R. Verba, K. O. Levchenko, S. Koraltan, K. Davídková, B. Budinská, Q. Wang, O. V. Dobrovolskiy, M. Urbánek, M. Lindner, T. Reimann, C. Dubs, C. Gonzalez-Ballestero, C. Abert, D. Suess, D. A. Bozhko, S. Knauer, and A. V. Chumak, \textit{npj Spintronics} \textbf{2}, 29 (2024).
\end{enumerate}

\clearpage

\end{document}